\documentclass[journal,draftcls,onecolumn,12pt,twoside]{IEEEtranTCOM}
\normalsize

\usepackage[hidelinks]{hyperref}
\usepackage{amsmath,amssymb,subcaption,amsthm,graphicx,caption}
\usepackage[font=scriptsize]{caption}
\usepackage{float}
\usepackage{bbm}
\begin{document}
%
\title{A Unified Statistical Model for Atmospheric Turbulence-Induced Fading in Orbital Angular Momentum Multiplexed FSO Systems  }

\author{El-Mehdi Amhoud,~\IEEEmembership{Member,~IEEE,} \mbox{Boon S. Ooi},~\IEEEmembership{Senior Member,~IEEE,}\\ and Mohamed-Slim Alouini,~\IEEEmembership{Fellow,~IEEE}

\thanks{El-Mehdi Amhoud was with  the Computer, Electrical, and Mathematical Science and Engineering Division, King Abdullah University of Science and Technology, Thuwal, Makkah Province, Saudi Arabia.  He is now with the School of Computer and Communication Sciences, Mohammed VI \mbox{Polytechnic} University, Morocco, email: elmehdi.amhoud@um6p.ma).
\newline
 Boon S. Ooi and Mohamed-Slim Alouini are with the Computer, Electrical, and Mathematical Science and	Engineering  Division, King Abdullah University of Science and Technology Thuwal, Makkah Province,
Saudi Arabia (e-mails: $\{\text{boon.ooi, slim.alouini}\}$@kaust.edu.sa).}
} 
\maketitle
\begin{abstract}
This paper proposes a unified statistical channel model to characterize the atmospheric turbulence induced distortions faced by orbital angular momentum (OAM) in free space optical (FSO) communication systems. In this channel model, the self-channel irradiance of OAM modes as well as crosstalk irradiances between different OAM modes are characterized by a Generalized Gamma distribution (GGD). The latter distribution is shown to provide an excellent match with simulated data for all regimes of atmospheric turbulence. Therefore, it can be used to overcome the computationally complex numerical simulations to model the propagation of OAM modes through atmospheric turbulent FSO channels. The GGD allows obtaining very simple tractable closed-form expressions for a variety of performance metrics. Indeed,  the average capacity, the bit-error rate, and the outage probability are derived  for FSO systems using single OAM mode transmission with direct detection.  Furthermore, we extend our study to FSO systems using OAM mode diversity to improve the performance. By using a maximum ratio combining (MRC) at the receiver,  the GGD is also shown to fit the simulated combined received optical powers. Finally, space-time (ST) coding is proposed to provide diversity and multiplexing gains, and the error probability is theoretically derived under the newly proposed generic model.
\end{abstract}

\IEEEpeerreviewmaketitle
\section{Introduction}
Space division multiplexing (SDM) technique allows to increase the capacity of a communication system by transmitting several independent data streams in parallel. In optical communications, SDM can be realized by sending different spatial modes in a multi-mode fiber or using many cores in a multi-core fiber \cite{Richardson}.  
In analogy to optical fiber transmission systems, orbital angular momentum (OAM)  is proposed  to transmit multiple signals  over  free space channels \cite{WillnerScience,Trichili}. This simultaneous transmission of information on OAM modes is possible thanks to the orthogonality property of OAM beams that allows propagation without interference between signals. A simple scheme of OAM FSO transmission consists in using only the intensity of OAM beams to carry the information. At the receiver, photo-detectors are used to detect  light intensity,  and the  system is operating in intensity-modulation direct-detection (IM/DD). This configuration has a low cost and is more feasible for practical deployment. On the other hand, coherent detection mostly used  with optical fibers  can also be used for OAM FSO systems. In coherent systems, the spectral efficiency can be further increased by using the two polarizations of light and higher order modulation formats. However, these benefits come at the cost of the high receiver expense and complexity that increases with the number of modes. 
By using OAM multiplexing as an additional degree of freedom to polarization and wavelength, coherent laboratory experiment has achieved more than 1 Pbit/s capacity using 26 modes \cite{1Pbits}.
\newline
In real life communication scenarios, transmitted OAM beams are subject to atmospheric turbulence induced distortions that if not properly addressed and compensated may deteriorate the entire FSO link \cite{Ren,li2014evaluation,anguita2008turbulence}. Several experimental and simulation works have studied the effect of atmospheric turbulence and different techniques to emulate turbulence have been proposed. Atmospheric turbulence can be emulated by using a rotatable phase plate that has a pseudo-random distribution that obeys Kolmogorov statistics. This technique allows to control the turbulence strength and was used in \cite{li2018ecoc}, at the transmitter for a 2-OAM transmission in a 100-m round trip FSO link. The same process was used in \cite{CodingFSOMIMO} to emulate turbulence in a 4-OAM modes transmission. An alternative to rotatable phase plate is spatial light modulator (SLM). In fact, the distorted phase hologram of a specific OAM mode after propagation  can be generated by computer and printed on a SLM. The latter is used at the receiver as a demodulating SLM that contains the effect of atmospheric turbulence \cite{sun2017crosstalk}. Besides, propagation of OAM modes through the turbulent atmosphere has also been subject to numerical simulation studies. This can be realized using the split-step Fourier propagation method where turbulence is induced along the propagation path \cite{anguita2008turbulence,belmonte2000feasibility,zhao2012aberration,liu2011investigation,amhoud2018oam}. This technique is very accurate and can simulate  the propagation of any OAM mode for different turbulence regimes. Nonetheless, it is computationally complex as it involves many multiplication operations of matrices having large dimensions. Therefore it is important to have a statistical model to describe the propagation of OAM beams in turbulent free space.
\par
In classical FSO systems employing only the Gaussian beam for transmission, several statistical models were proposed for irradiance fluctuations. For example, a Log-Normal distribution can be used to model weak turbulence variations \cite{obukhov1953effect}. For moderate, to strong turbulence, a Double-Weibull stochastic model was proposed in \cite{chatzidiamantis2010new}. Moreover, the Gamma-Gamma distribution is a widely accepted model used for all levels of turbulence \cite{al2001mathematical}.  For all the previous models, closed-form expressions for performance metrics such as the average capacity, the outage probability, and the bit-error rate (BER) were derived.
In OAM FSO systems, to the best of our knowledge, the only proposed statistical channel models for turbulence-induced fading were given in \cite{Anguita,Funes}. In this models, the self-channel irradiance of OAM modes was shown to obey a \mbox{Johnson $S_B$} distribution whereas the crosstalk between OAM modes follows an Exponential distribution. Knowing that the statistical properties of the Johnson $S_B$ distribution are analytically intractable, it is not possible to extract closed-form expressions for performance metrics. Moreover, the lack of a unified turbulence model makes it difficult to assess the effect of the interference on the performance of OAM modes. In \cite{Alfowzan}, the interference fading between OAM modes was assumed to be   part of the added Gaussian noise. This consideration has allowed deriving a theoretical BER expression but was unfortunately  not corresponding to the numerically simulated BER.
\newline
In this work, we show that the Generalized Gamma distribution (GGD) can efficiently  model the turbulence induced self-channel fading as well as the crosstalk between OAM modes. The GGD  gives a perfect fit for the weak, moderate,  and strong turbulence regimes. Closed-form expressions for the outage probability, the average capacity, and the BER are derived for single-input single-output (SISO) transmission using only one OAM mode. Afterward, we consider a multiple-input multiple-output (MIMO) system using several OAM modes to provide diversity gain and hence improve the performance. Furthermore, space-time (ST) coding is proposed to achieve diversity and multiplexing gains, and the corresponding performance is analyzed.
\newline
The remainder of this paper is organized as follows. In Section II, we describe the OAM modes generation and propagation in turbulence, and then we  present the GGD and its main statistical parameters. In Section III, we compare our proposed GGD model to numerical simulations of atmospheric turbulence fading and also to the Johnson $S_B$ and Exponential models. In Section IV, we derive closed-form expressions for performance metrics using the GGD model and compare the fitting goodness with numerical simulations. Afterward,  spatial diversity using several OAM modes is studied and shown to give better performance than SISO links in section V.  In section VI, ST coding is proposed to achieve  full diversity and multiplexing gains  and the theoretical error probability is derived. Finally,  the conclusion is given in Section VII.
\section{Channel Model }
\subsection{Orbital Angular Momentum Multiplexing}
By definition, an electromagnetic wave carrying an OAM of $m\hbar$ per photon is a wave possessing an azimuthally varying phase term $\exp(im\phi)$, where $m$ is an unbounded integer known as the topological charge, $\phi$ is the azimuth and $\hbar$ is the reduced Planck constant \cite{AllenPRA92}.
To realize OAM multiplexing, single and superposition of orthogonal beams that have a well-defined vorticity can be used including  Hermite-Gaussian (HG) beams \cite{Soares}, Ince-Gaussian beams \cite{Ince}, Bessel-Gaussian beams  and Laguerre-Gauss (LG) beams \cite{Doster}. In this work, we are interested in OAM  derived from LG modes. The spatial distribution of the LG beams is given by \cite{Doster}:
\begin{multline}
\begin{aligned}
& u\left ( r,\phi ,z \right )= \sqrt{\frac{2p!}{\pi(p+\left | m \right |)!}}\frac{1}{w(z)}\left [ \frac{r\sqrt{2}}{w(z)} \right ]^{\left | m \right |} L^{\left | m \right |}_p\left ( \frac{2r^2}{w^2(z)} \right ) 
 \times \exp\left (\frac{-r^2}{w^2(z)}  \right )   \\
&\times \exp\left ( \frac{-ikr^2z}{2(z^2+z^2_R)} \right ) \times \exp \left ( i(2p + \left | m \right | + 1) \tan^{-1}\left ( z/z_R \right ) \right )  \exp\left ( -im\phi \right ),
\end{aligned}
\end{multline}
where $r$ refers to  the radial distance. In \cite{Doster},\mbox{ $w\left ( z \right )=w_0\sqrt{\left ( 1+\left ( z/z_R \right )^2 \right )}$} is the beam radius at the distance $z$, where $w_0$ is the beam waist of   the Gaussian beam, $z_R=\pi w^2_0/\lambda$ is the Rayleigh range, and $\lambda$ the optical carrier. In \cite{Doster}, $k=2\pi/ \lambda$ is the propagation constant, $ L^m_p(\cdot )$ is the generalized Laguerre polynomial, where $p$ and $m$ represent the radial and angular mode numbers. OAM modes correspond to the subset of LG modes having $p=0$ and $m\neq 0$.
\newline
OAM modes generation can  be realized in practice by different techniques such as spiral phase plates (SPP) \cite{Massari},  and spatial light modulators (SLM) \cite{Ohtake}. An SPP is an optical device that has a spiral staircases form. This geometric shape allows to transform a Gaussian incident beam into a beam with a helical phase front. SPP is an efficient and stable manner to generate OAM beams. However a single SPP allows the generation of only one OAM beam. On the other hand, SLM allows to dynamically generate OAM beams. This is done thanks to computer generated holograms that are printed on the SLM. 
\begin{figure}
	\centering
	\includegraphics[width=\columnwidth,height=4cm]{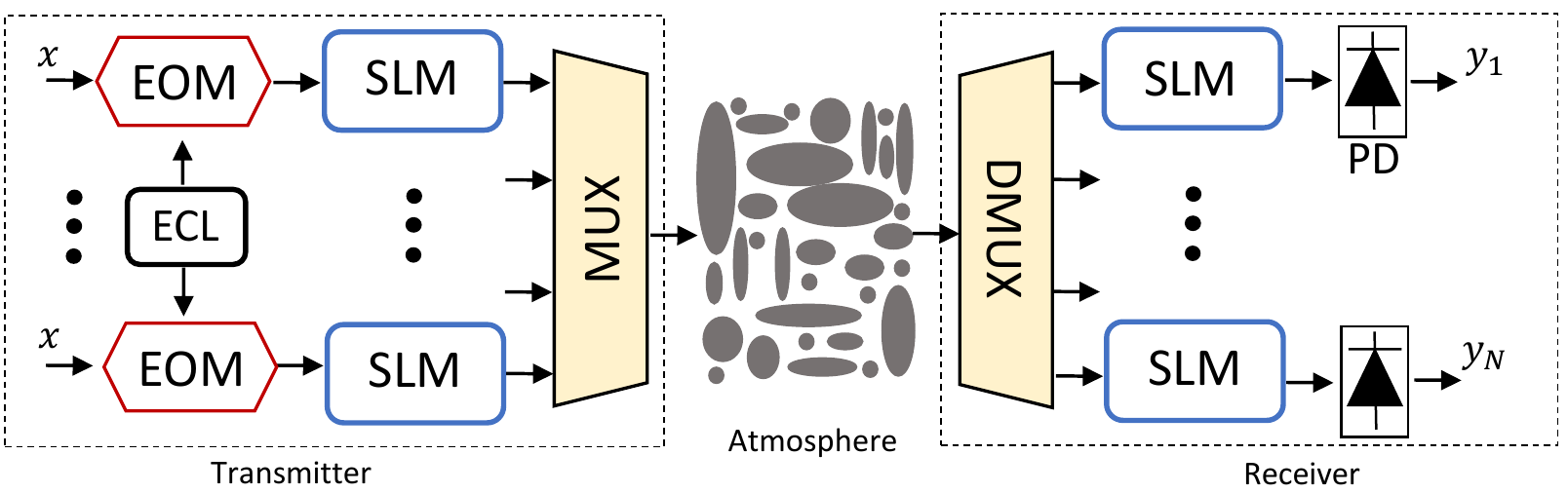}
	\caption{FSO transmitter and receiver front ends: ECL: External cavity laser, EOM: Electro-optical modulator, SLM: Spatial light modulator, MUX/DMUX: Multiplexer/Demultiplexer, and PD: Photodiode.}
	\label{channel}
\end{figure}
 A  transmission scheme of OAM FSO system is represented in Fig. \ref{channel}. The electro-optical modulator (EOM) modulates the signal of the external laser source (ECL). The output optical signal of the EOM is still a Gaussian beam. Thanks to printed  holograms,  SLMs create the desired OAM modes, then signals are multiplexed and sent into the  FSO channel. At the receiver signals are demultiplexed and then inverse operations on the SLMs are performed to convert back signals to Gaussian beams that are detected using the photodiodes.
The vorticity of OAM  beams propagating in free space without atmospheric turbulence is preserved. Consequently, OAM beams maintain orthogonality as they propagate which can be described by \cite{Ren,Huang}:
\begin{equation}
\int u_m(\boldsymbol{r},z) u_n^\ast (\boldsymbol{r},z)d\boldsymbol{r}=\left\{\begin{matrix}
1~,~~\text{if}~ m=n\\ 
0~,~~\text{if}~ m\neq n
\end{matrix}\right.,
\end{equation} 
where $u_m (\boldsymbol{r},z)$ refers to the normalized field distribution of OAM mode of order $m$ at the distance  $z$ and $\boldsymbol{r}$ refers to the radial position vector. 
\subsection{OAM Propagation in Turbulence}
The propagation of OAM modes can be affected by atmospheric turbulence induced distortions \cite{Ren,CodingFSOMIMO,Huang,TurbulenceEffects}. Atmospheric turbulence is  caused by pressure and temperature fluctuations in the atmosphere which result in a random behavior of the atmospheric refractive index \cite{Rodenburg}. These effects make the atmospheric refractive index to be spatially dependent which distorts the propagating light beams and causes the break of their orthogonality. The  spatial fluctuations of the atmospheric refractive index obey a modified Kolmogorov spectrum given by \cite{andrews1992analytical}: 
\begin{eqnarray}
\Phi (\kappa )=0.033C^2_n\frac{\exp(-\kappa ^2/\kappa_l^2)}{(\kappa ^2+\kappa_0 ^2)^{11/6}}f(\kappa,\kappa_l),
\end{eqnarray}
where $f(\kappa,\kappa_l)=[1+1.802(\kappa/\kappa_l)-0.254(\kappa/\kappa_l)^{7/6}] $. $C^2_n$ is the refractive index structure parameter. $\kappa_0=2\pi/L_0$, with $L_0$ is the outer scale of the turbulence. $\kappa_l=\frac{3.3}{l_0}$, with $l_0$ is the inner scale of the turbulence. To emulate atmospheric turbulence, random phase screens are placed along the FSO channel. 
The strength of the turbulence in a FSO channel is given by the Rytov variance defined as $\sigma _R^2=1.23C^2_n(2\pi/\lambda )^{7/6}z^{11/6}$.  Weak atmospheric turbulence is usually associated with $\sigma _R^2<1$, whereas strong turbulence  refers to $\sigma _R^2>1$ \cite{anguita2008turbulence,Huang}. Saturation fluctuations of atmospheric turbulence occurs when $\sigma _R^2>>1$ \cite{chatzidiamantis2010new,katsilieris2017accurate}. In our work, we define three regimes of atmospheric turbulence depending on the Rytov variance as follows \cite{katsilieris2017accurate}:
\begin{equation}
\begin{matrix}
\sigma _R^2\leq 0.3 & \text{weak~turbulence}\\ 
0.3< \sigma _R^2\leq 5 & \text{moderate~to strong~turbulence}\\ 
\sigma _R^2 > 5 & \text{saturation regime.} 
\end{matrix}
\label{ATregimes}
\end{equation}
Due to atmospheric turbulence, the signal initially launched on an OAM mode $m$ will spread to other OAM modes. The leakage of power is more important for adjacent modes to the mode $m$ and tends to zero for further OAM modes from mode $m$ as shown in \cite{Anguita,Huang}.
\par
Given that a set of $\mathcal{M}$ OAM modes can be launched at the transmitter and $\mathcal{N}$ OAM modes can be detected at the receiver. The transmission of OAM modes in turbulent FSO channel can be described by a multiple-input multiple-output system where the received signal on  the $n^{\text{th}}\in \mathcal{N}$ OAM mode is given by:  
\begin{eqnarray}
y_n=x\eta\sum_{m \in \mathcal{M}}^{}I_{mn}+v_n,
\label{MIMO}
\end{eqnarray}   
where $x\in \left \{ 0, 1 \right \}$ is the transmitted information bit. $\eta$ is the optical-to-electrical conversion coefficient. $v_n$ is an additive white Gaussian noise with zero mean and variance equal to $N_0$. $I_{mn}$ represents the fraction of power received on the analyzing field of OAM mode of topological charge $n$ when the  transmitted power is carried by  OAM mode with topological charge $m$. $I_{mn}$ is given by the square of the scalar product between the fields of the two modes as  \cite{anguita2008turbulence,mehrpoor2015free}: 
\begin{equation}
I_{mn}= \left | \int u_m(\boldsymbol{r},z) u_n^\ast (\boldsymbol{r},z)d\boldsymbol{r} \right |^2.
\label{Imn}
\end{equation}
$I_{mn}$ was shown to have an Exponential distribution for $m \neq n$ in \cite{Anguita}, and a Johnson $S_B$ distribution for  $m = n$ \cite{Funes}. In the next section we introduce the GGD and present some of its useful statistical properties that we will use latter for performance metrics derivation.
\subsection{Generalized Gamma Distribution}
The GGD was proposed by Stacy in \cite{stacy1962generalization}. Its probability density function (PDF) is given by:
\begin{equation}
f_I(I;a,b,c)=c\frac{I^{ac-1}}{b^{ac}}\frac{\exp(-(I/b)^c)}{\Gamma(a)},
\label{PDF_GGD}
\end{equation}
where $a$ and $c$ are the shape parameters, $b$  is the scale parameter of the GGD, and $\Gamma(\cdot)$ denotes the Gamma function. The GGD is a very flexible distribution that includes several well-known distributions. For $c=1$, Eq. (\ref{PDF_GGD}) reduces to the Gamma distribution. For $a=1$, Eq. (\ref{PDF_GGD}) becomes a Weibull distribution. When $a=c=1$, Eq. (\ref{PDF_GGD}) corresponds to the Exponential distribution. Furthermore, as $a\rightarrow \infty $, Eq. (\ref{PDF_GGD}) leads to a Log-Normal distribution \cite{khodabina2010some}. By using \cite[Eq. (2.9.4)]{kilbas2004h} followed by \cite[Eq. (2.1.4)]{kilbas2004h} and \cite[Eq. (2.9.1)]{kilbas2004h}, we can rewrite the GGD in terms of the Meijer's G function as:

\begin{equation}
f_I(I;a,b,c)=\frac{c}{I\Gamma(a) }\text{G}_{0,1}^{1,0}\left [ \left ( \frac{I}{b} \right )^c  \left|\begin{matrix}
-\\ 
a
\end{matrix}\right. \right ].
\label{PDF_GGD_Meijer}
\end{equation}
The cumulative distribution function (CDF) of $I$ can be obtained using the definition of the Meijer's G function in  \cite[Eq. (2.9.1)]{kilbas2004h}:
\begin{equation}
F_{I}(I;a,b,c)=\frac{1}{\Gamma(a) }\text{G}_{1,1}^{1,2}\left [ \left ( \frac{I}{b} \right )^c  \left|\begin{matrix}
1\\ 
a,0
\end{matrix}\right. \right ].
\label{CDF_I}
\end{equation}
The $k^{\text{th}}$ moment of the GGD defined as $\mathbb{E}\left [ I^k \right ]= \int_{0}^{\infty }I^kf_I(I)dI$ is given by \cite[Eq. (2)]{khodabina2010some}:
\begin{equation}
\mathbb{E}\left [ I^k \right ]=b^k\frac{\Gamma(a+\frac{k}{c})}{\Gamma(a) },
\end{equation}
where $\mathbb{E}[\cdot]$ is the expectation operator.
The moment generating function (MGF) of the GGD can be derived in terms of the Fox's H function as \cite[Eq. (27)]{zedini2019unified}: 
\begin{equation}
\textup{M}_{I}\left ( I \right ) = \frac{1}{\Gamma(a) }\text{H}_{1,1}^{1,1}\left [ -Ib \left|\begin{matrix}
(1-a,1/c)\\ 
(0,1)\end{matrix}\right. \right ].
\label{MGF}
\end{equation}
To evaluate the Fox's H function, an efficient MATHEMATICA implementation can be found in \cite[Appendix A]{yilmaz2009product}. A simpler expression for the MGF is given in the form of an infinite sum as \cite[Eq. (20)]{stacy1962generalization}:
\begin{equation}
\textup{M}_{I}\left ( I \right ) = \frac{1}{\Gamma(a) }\sum_{k=0}^{\infty }\Gamma\left ( \frac{k}{c}+a \right )\frac{\left ( bI \right )^k}{k!}.
\label{MGF2}
\end{equation} 

\section{GGD Fitting Comparison with simulations}
\begin{figure*}
	\centering
	\includegraphics[width=0.9\textwidth,height=\textheight ]{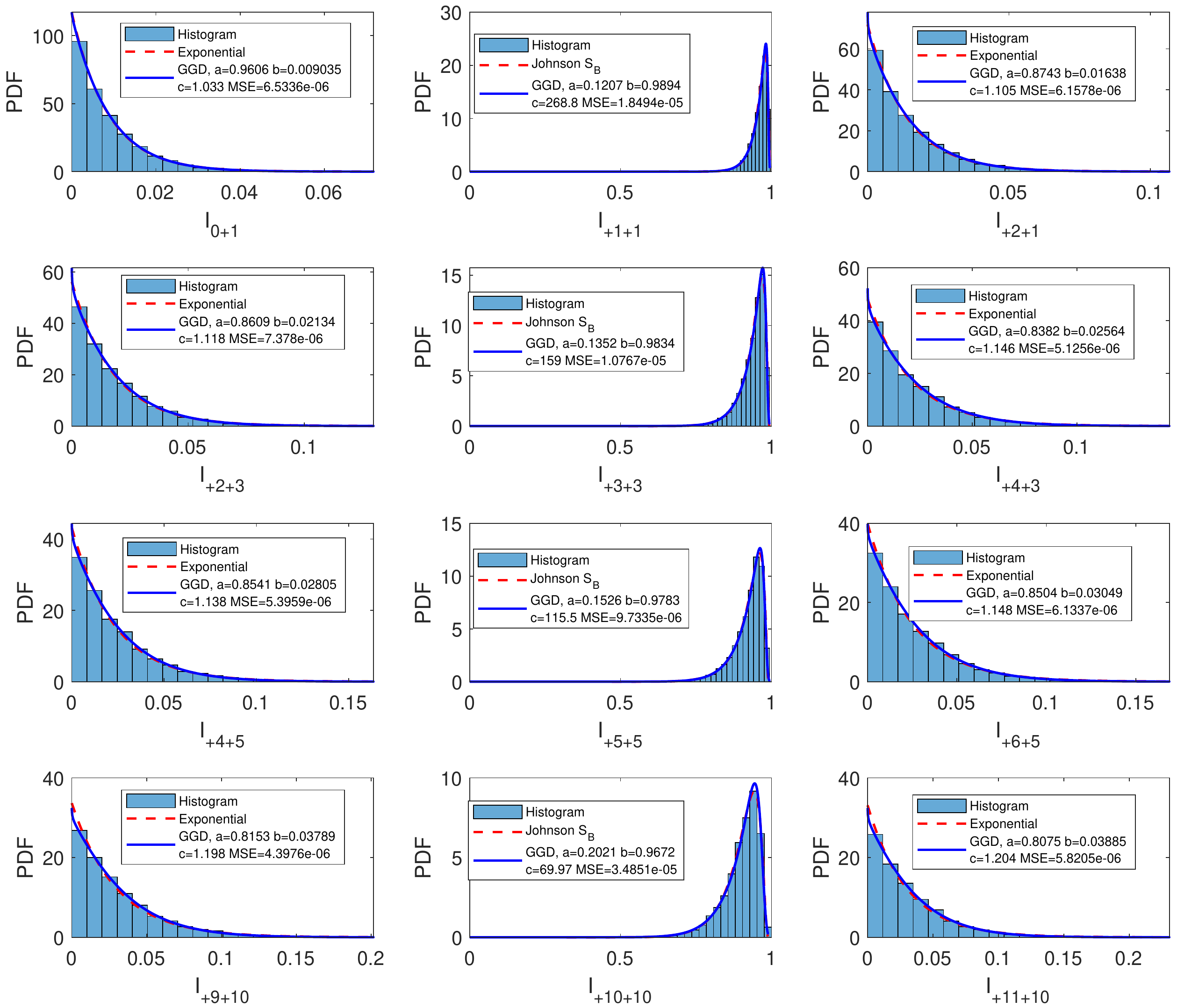}
	\caption{Distribution of the self-channel irradiance for OAM modes in the set $\mathcal{M}=\left \{ +1, +3, +5, +10 \right \}$ and  crosstalk irradiances from adjacent OAM modes in the weak turbulence regime $C^2_n=5\times 10^{-15}$, $\sigma _R^2=0.2$.}
	\label{weak_turbulence_dist}
\end{figure*}

\begin{figure*}
	\centering
	\includegraphics[width=0.9\textwidth,height=\textheight ]{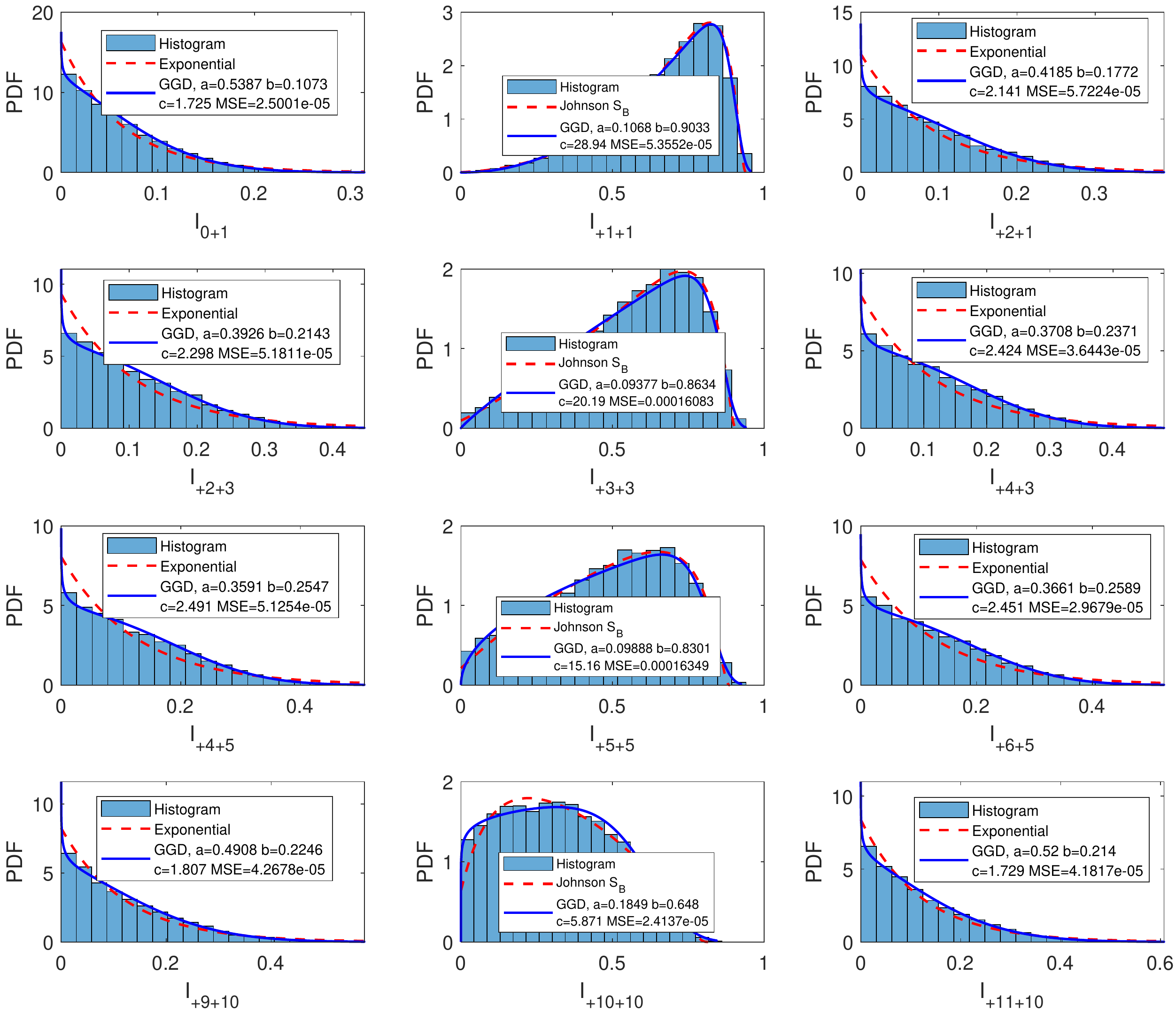}
	\caption{Distribution of the self-channel irradiance for OAM modes in the set $\mathcal{M}=\left \{ +1, +3, +5, +10 \right \}$ and  crosstalk irradiances from adjacent OAM modes in the moderate to strong turbulence regime $C^2_n=7\times 10^{-14}$, $\sigma _R^2=2.80$.}
	\label{moderate_turbulence_dist}
\end{figure*}

\begin{figure*}
	\centering
	\includegraphics[width=0.9\textwidth,height=\textheight ]{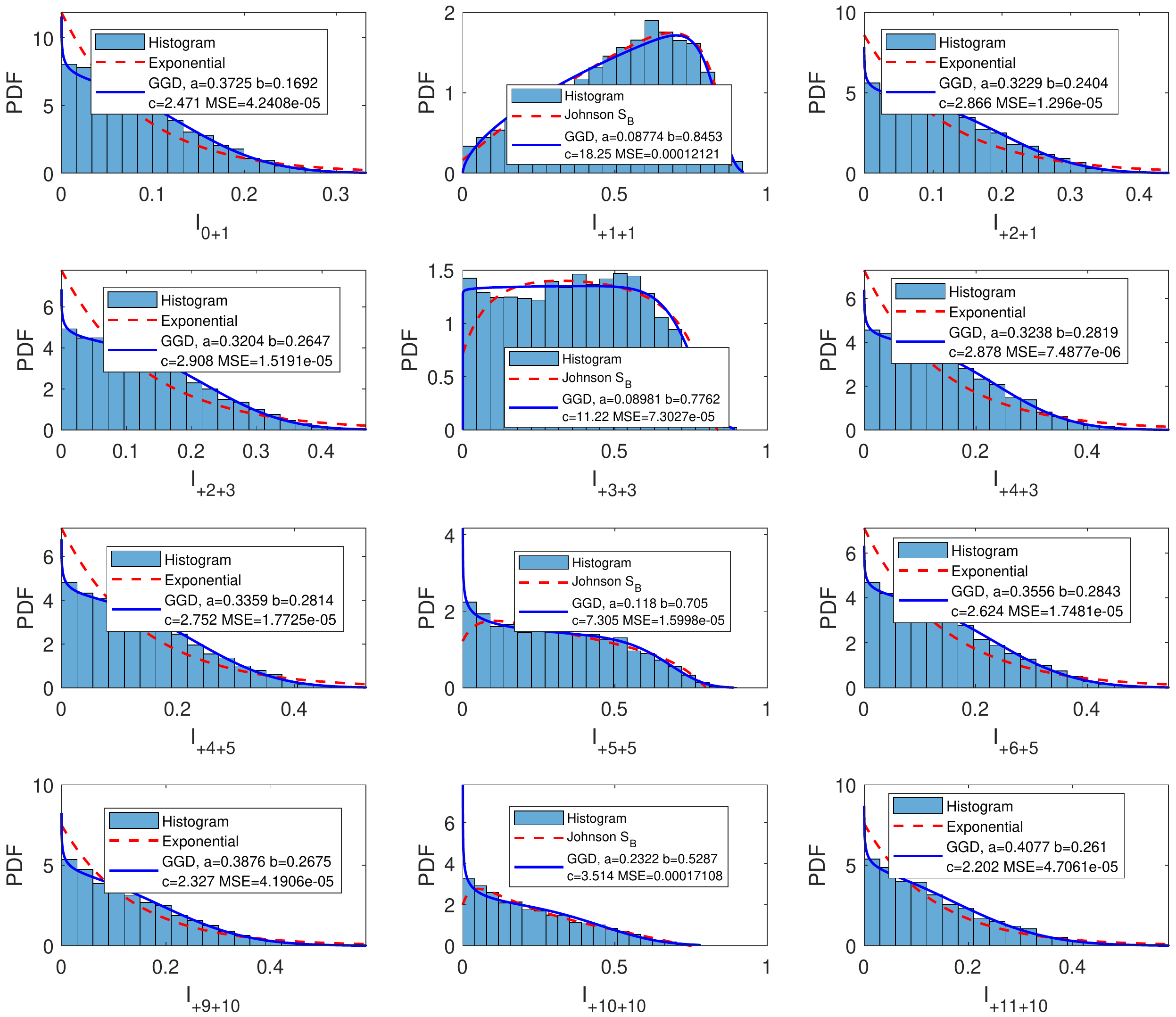}
	\caption{Distribution of the self-channel irradiance for OAM modes in the set $\mathcal{M}=\left \{ +1, +3, +5, +10 \right \}$ and  crosstalk irradiances from adjacent OAM modes in the saturation regime $C^2_n=3\times 10^{-13}$, $\sigma _R^2=12$.}
	\label{saturation_turbulence_dist}
\end{figure*}

In this section, we evaluate the suitability of the proposed GGD to fit the simulation data.
As we are working in an IM/DD setup, our Monte Carlo simulations of OAM propagation through the turbulent atmosphere are performed at wavelength  $\lambda= 850~$nm. The beam waist  at the transmitter is $\omega_0=1.6$~cm. On the other hand, the optical receiver is assumed to be large enough to collect all the power on the OAM modes. This condition is satisfied when the receiver diameter  is  $d_{Rx}=2w_z\sqrt{m_{\text{max}}}$, where $w_z$ is the radius of the Gaussian beam and $m_{\text{max}}$ is the highest OAM topological charge used in the system. 
 To create the desired OAM modes, SLMs having \mbox{$512 \times 512$} pixels are used. The propagation distance is set to $z=1~$km, the inner and outer scales of turbulence are set to $l_0=5~$mm and $L_0=20~$m, respectively. Atmospheric turbulence is emulated by placing $20$ random phase screens each  $50~$m along the propagation path. Each phase screen is evaluated  as the Fourier transform of a complex random distribution with zero mean and variance equal to  $\left ( \frac{2\pi}{N\Delta x} \right )^2\Phi (\kappa )$, where $N=512$ is the array length and $\Delta x=1.36~$mm is the grid spacing assumed to be equal in both dimensions $x$ and $y$.  Propagation through the turbulent atmosphere is simulated using the commonly used split-step Fourier method \cite{belmonte2000feasibility}. 
 \newline
 Let $I_{mn}^1,...,I_{mn}^N$ be the set of irradiance realizations for a particular $m$ and $n$.  $N$ is the number of samples, where 
 for each $I_{mn}$ we have simulated $N=10^6$ realizations. We use a maximum-likelihood (ML) estimation in order to estimate the parameters $a$, $b$, and $c$ of $I_{mn}$. The ML estimator computes the loglikelihood function given by \cite{wong1993simultaneously}:
\begin{multline}
\begin{aligned}
 &L\left (I_{mn}^i; a,b,c \right )=N\log\left ( c/\left ( b^{ac} \Gamma\left ( a \right ) \right ) \right ) \\
 &+ \left ( ac-1 \right )\sum_{i=1}^{N}\log(I_{mn}^i)- b^{-c}\sum_{i=1}^{N}\left ( I_{mn}^i \right )^c.
 \label{ML}
\end{aligned}
\end{multline}
The estimated parameters $a$, $b$, and $c$ represent the values for which the previous function is maximized and are giving by \cite{wong1993simultaneously,song2008globally}:
\begin{equation}
a =\left [ c \left ( \frac{\sum_{i=1}^{N}\left ( I_{mn}^i \right )^c \log\left ( I_{mn}^i \right )}{\sum_{i=1}^{N}\left ( I_{mn}^i \right )^c}-\frac{\sum_{i=1}^{N}\log\left ( I_{mn}^i \right ) }{N} \right )\right ]^{-1}
\end{equation}

 \begin{equation}
b= \left ( \frac{\sum_{i=1}^{N}\left ( I_{mn}^i \right )^c}{Na} \right )^{1/c}
 \end{equation}

 \begin{equation}
\frac{c}{N}\sum_{i=1}^{N}\log(I_{mn}^i)-\log\sum_{i=1}^{N}\left ( I_{mn}^i \right )^c+\log\left ( Na \right )-\psi _0\left ( a \right ),
 \end{equation}
where $\psi _0$ denotes the digamma function. In our simulations, we have used the ML implementation given in the wafo toolbox \cite{wafo_toolbox} . 
 \newline
 To validate the goodness of the  proposed GGD fit, we used the mean square error (MSE) test. The latter can give a measure on the accuracy of the GGD fitting to the simulated data. The MSE estimator is defined as:
\begin{equation}
\text{MSE}=\frac{1}{N}\sum_{k=1}^{N}\left ( \hat{F}^k(I_{mn})-F^k_{I_{mn}}(I_{mn}) \right )^2,
\end{equation}
where $\hat{F}^k(I_{mn})$ is the empirical distribution function of $I_{mn}$ and
$F^k_{I_{mn}}(I_{mn})$  is the cumulative distribution function of $I_{mn}$ computed using the estimated GGD parameters of  $I_{mn}$.
\newline
In Figs. \ref{weak_turbulence_dist}, \ref{moderate_turbulence_dist}, and \ref{saturation_turbulence_dist}, we plot the distributions of $I_{mn}$ for  $m\in \mathcal{M}=\left \{ +1, +3, +5, +10 \right \}$ and $n\in \mathcal{N}=\left \{ m-1,m,m+1 \right \}$ for different atmospheric turbulence regimes. Also for each distribution, a GGD fitting along with a Johnson $S_B$ or Exponential distribution fitting are plotted. 
\newline
In Fig. \ref{weak_turbulence_dist}, a weak turbulence regime given by $C^2_n=5\times 10^{-15}$ and $\sigma _R^2=0.2$ is considered. The center column of the figure shows the obtained results for the self-channel irradiance (i.e., the distribution of the power transmitted on OAM mode $m$ and detected on the same OAM mode $n=m$). The obtained distributions have a perfect correspondence with the distributions obtained in \cite{Anguita}. The histograms are left skewed and very close to the value of $1$. This is explained by the fact that almost all power transmitted on a specific OAM mode remains in this mode when atmospheric turbulence is weak. Furthermore, we notice that the GGD gives a perfect fit with the obtained histograms and very low values of the MSE were achieved.  In addition, the left and right columns of Fig. \ref{weak_turbulence_dist} shows the distribution of the crosstalk of the adjacent OAM modes to the transmitted OAM mode $m$. We notice the Exponential shape of the crosstalk histograms which are perfectly fitted by the GGD, and achieving an MSE always less than $4 \times 10^{-6}$. 
\newline
In Fig. \ref{moderate_turbulence_dist}, a moderate to strong atmospheric turbulence regime with $C^2_n=7\times 10^{-14}$ and $\sigma _R^2=2.80$ is considered. From the center column of Fig. \ref{moderate_turbulence_dist}, we notice that the effect of turbulence is remarkable on the distributions of $I_{mm}$ which are no longer tight and close to 1. In this regime of turbulence, the GGD is also capable of fitting the simulated data histograms with an MSE less than $1.6 \times 10^{-4}$. For OAM mode $m=+10$, the GGD provides a better fit than the Johnson $S_B$. In addition, the crosstalk histograms  are perfectly fitted by the GGD and its fitting is also much better than the Exponential fit. Lastly, for the saturation regime, we have considered $C^2_n=3\times 10^{-13}$ and  $\sigma _R^2=12$. As for the previous two cases, the GGD gives an excellent  fit for all the obtained distributions and achieves a very low MSE. Moreover, for the crosstalk distributions, we notice from Fig. \ref{saturation_turbulence_dist} that the GGD gives a better fit than the Exponential fit. Consequently, all the previous results indicate that the GGD is a very suitable candidate to model the irradiance distributions of OAM FSO transmission systems in the presence of atmospheric turbulence. Furthermore, we also notice from Figs. \ref{weak_turbulence_dist}, \ref{moderate_turbulence_dist}, and \ref{saturation_turbulence_dist} that the parameters $a$, $b$, and $c$ of the GGD depend on the OAM mode as well as the atmospheric turbulence strength. This situation is more complex than the classical FSO systems where the parameters of the irradiance of the Gaussian beam modeled as a Gamma-Gamma distribution could be directly derived from the inner and outer scale of the atmospheric turbulence \cite{al2001mathematical}. Consequently, the computation of the parameters of the GGD in OAM FSO systems as a function of the OAM mode order and the turbulence strength is subject of future works. 
\section{Performance Analysis of SISO OAM Links}
In this section, we consider that only one OAM mode is transmitted and detected in a system operating in intensity modulation and direct detection. For  notational convenience, we omit the subscript $m$ and refer to $I_{mm}$ by $I$.  The instantaneous  electrical signal-to-noise ratio (SNR) is defined as $\gamma=\left ( \eta I \right )^2/N_0$. The average electrical SNR is defined as $\mu=\left ( \eta \mathbb{E}\left \{  I\right \} \right )^2/N_0$. After a simple power transformation of the variable $I$ in Eq. (\ref{PDF_GGD_Meijer}), the PDF of the instantaneous SNR can be written as:
\begin{equation}
f_{\gamma}(\gamma;a,b,c)=\frac{c}{2\gamma \Gamma(a) }\text{G}_{0,1}^{1,0}\left [ \left ( \frac{\gamma}{b^2 \mu} \right )^{c/2}  \left|\begin{matrix}
-\\ 
a
\end{matrix}\right. \right ].
\label{PDF_SNR}
\end{equation}
The CDF of the  SNR defined as: $F_{\gamma}\left ( \gamma \right )=\int_{0}^{\infty }f_{\gamma}\left ( \gamma \right )d\gamma$ can be obtained using the Meijer's G function definition in \cite[Eq. (2.9.1)]{kilbas2004h}:
\begin{equation}
F_{\gamma}(\gamma;a,b,c)=\frac{1}{\Gamma(a) }\text{G}_{1,2}^{1,1}\left [ \left ( \frac{\gamma}{b^2 \mu} \right )^{c/2}  \left|\begin{matrix}
1\\ 
a,0
\end{matrix}\right. \right ].
\label{CDF_SNR}
\end{equation}
Now that we have defined the PDF and CDF of the  SNR, we can proceed to the performance metrics derivation.
\subsection{Ergodic Capacity}
The ergodic capacity is defined as \cite[Eq. (26)]{lapidoth2009capacity}, \cite{chaaban2016free}:
\begin{equation}
\text{C}_{erg}=\int_{0}^{\infty }\ln\left ( 1+\frac{e}{2\pi }\gamma  \right )f_{\gamma}(\gamma)d\gamma.
\label{Capacity}
\end{equation}
A closed-form expression for the ergodic capacity can be derived in terms of the Fox's H function as \cite[Eq. (31)]{zedini2019unified}:
\begin{equation}
\text{C}_{erg}=\frac{1}{\Gamma(a) }\text{H}_{2,3}^{3,1}\left [ \left ( \frac{2 \pi }{b^2 e \mu} \right )^{c/2}  \left|\begin{matrix}
(0,c/2)(1,1)\\ 
(a,1)(0,1)(0,c/2)
\end{matrix}\right. \right ].
\label{Cerg}
\end{equation}

\subsection{Outage Probability}
The outage probability $\text{P}_{out}$ is the probability that the instantaneous SNR falls bellow an SNR threshold $\gamma_{th}$ that ensures a reliable communication. It is given by the CDF of $\gamma$ evaluated at the value $\gamma_{th}$ as:
\begin{equation}
\text{P}_{out}=\text{Pr}\left ( \gamma < \gamma_{th} \right ) =F_{\gamma}\left ( \gamma_{th} \right ).
\label{Pout}
\end{equation}
\subsection{Average BER}
The BER expression for IM/DD FSO systems can be obtained using \cite[Eq. (19)]{zedini2017dual}:
\begin{equation}
\text{P}_e=\frac{1}{2\Gamma(\frac{1}{2})}\int_{0}^{\infty }\Gamma\left ( \frac{1}{2}, \frac{\gamma}{4}\right )f_{\gamma}(\gamma)d\gamma,
\end{equation}
where $\Gamma(\cdot ,\cdot )$ is the upper incomplete Gamma function. The previous equation can also be derived in closed-form expression given by the Fox's H function as \cite[Eq. (27)]{zedini2019unified}:
\begin{equation}
\text{P}_e=\frac{1}{\Gamma(a) }\text{H}_{2,2}^{1,2}\left [ \left ( \frac{4 }{b^2  \mu} \right )^{c/2}  \left|\begin{matrix}
(1,1)(\frac{1}{2},\frac{c}{2})\\ 
(a,1)(0,1)\end{matrix}\right. \right ].
\label{BER}
\end{equation}
\subsection{Numerical Simulations Evaluation }
\begin{figure}[h]
	\centering
	\begin{subfigure}[b]{0.45\columnwidth} 
		\centering
		\includegraphics[width=\columnwidth, height=6cm]{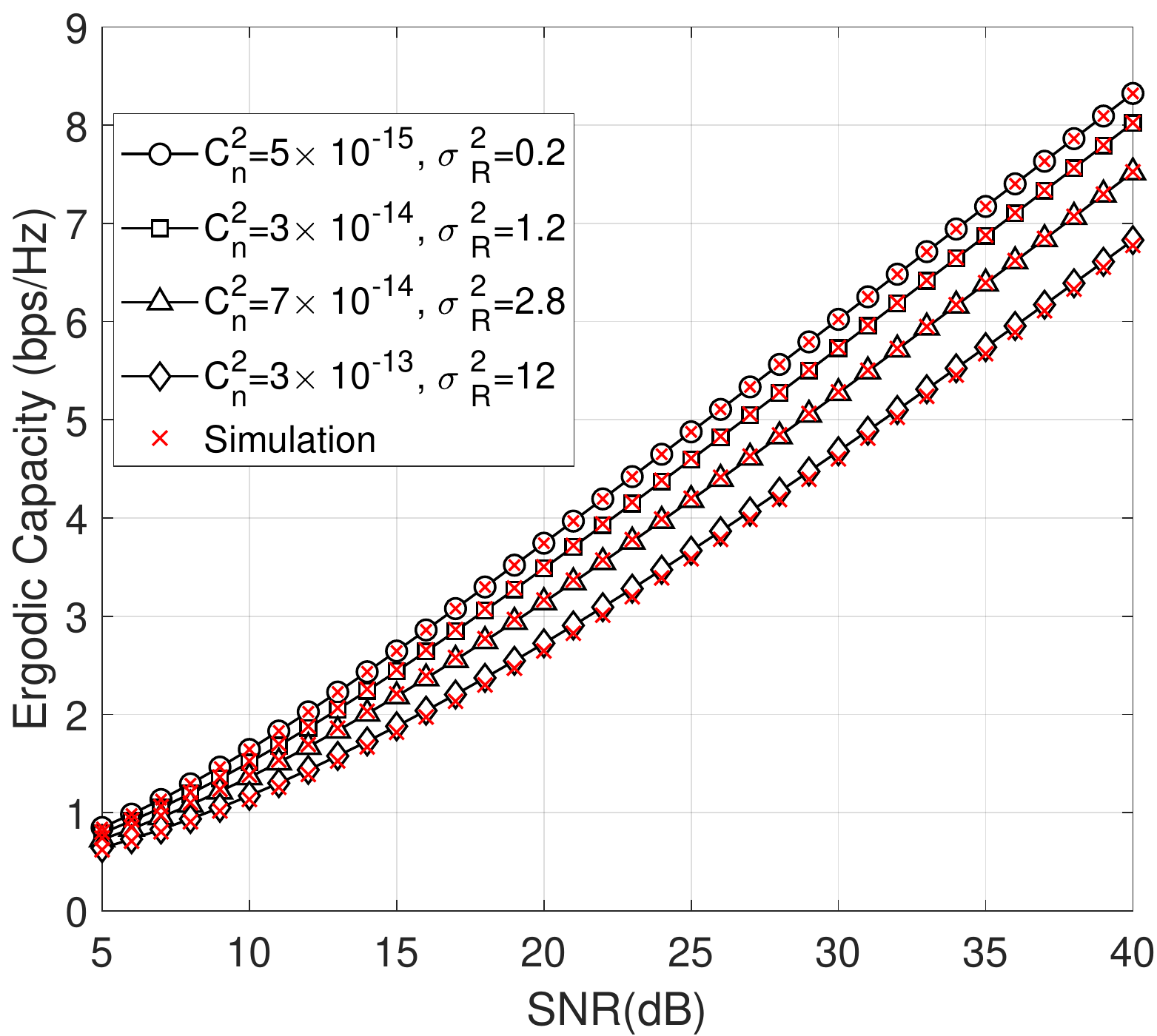}
		\caption{}
	\end{subfigure}~
	\begin{subfigure}[b]{0.45\columnwidth} 
		\includegraphics[width=\columnwidth, height=6cm]{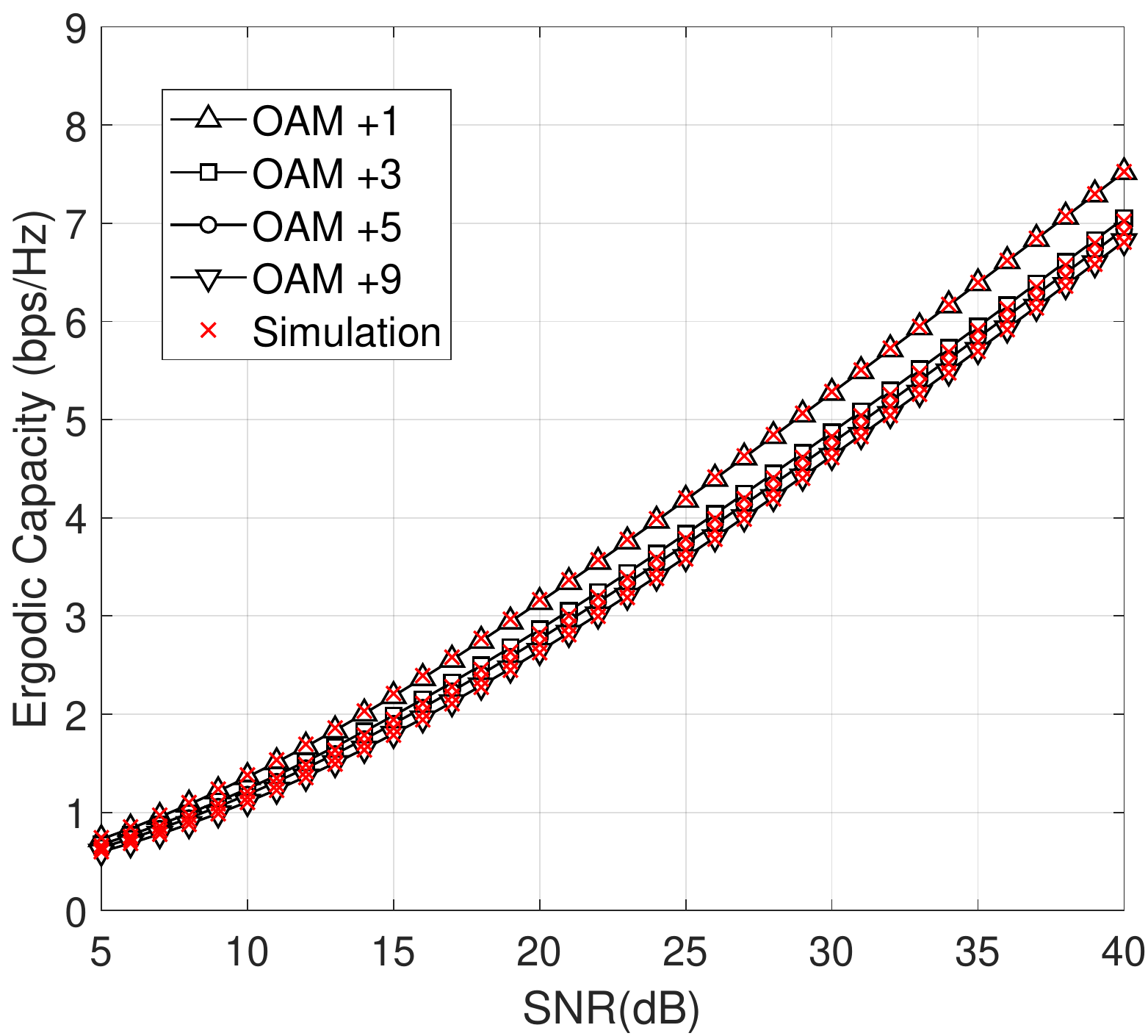}
		\caption{}
	\end{subfigure}
	\caption{(a): Ergodic capacity vs SNR for OAM mode $m=+1$ for different levels of atmospheric turbulence. (b): Ergodic capacity vs SNR for different OAM modes with $C^2_n=4\times 10^{-14}$ and  $\sigma _R^2=1.60$.}
	\label{CapacitiesErg}
\end{figure}

\begin{figure}[h]
	\centering
	\begin{subfigure}[b]{0.45\columnwidth} 
		\centering
		\includegraphics[width=\columnwidth, height=6.5cm]{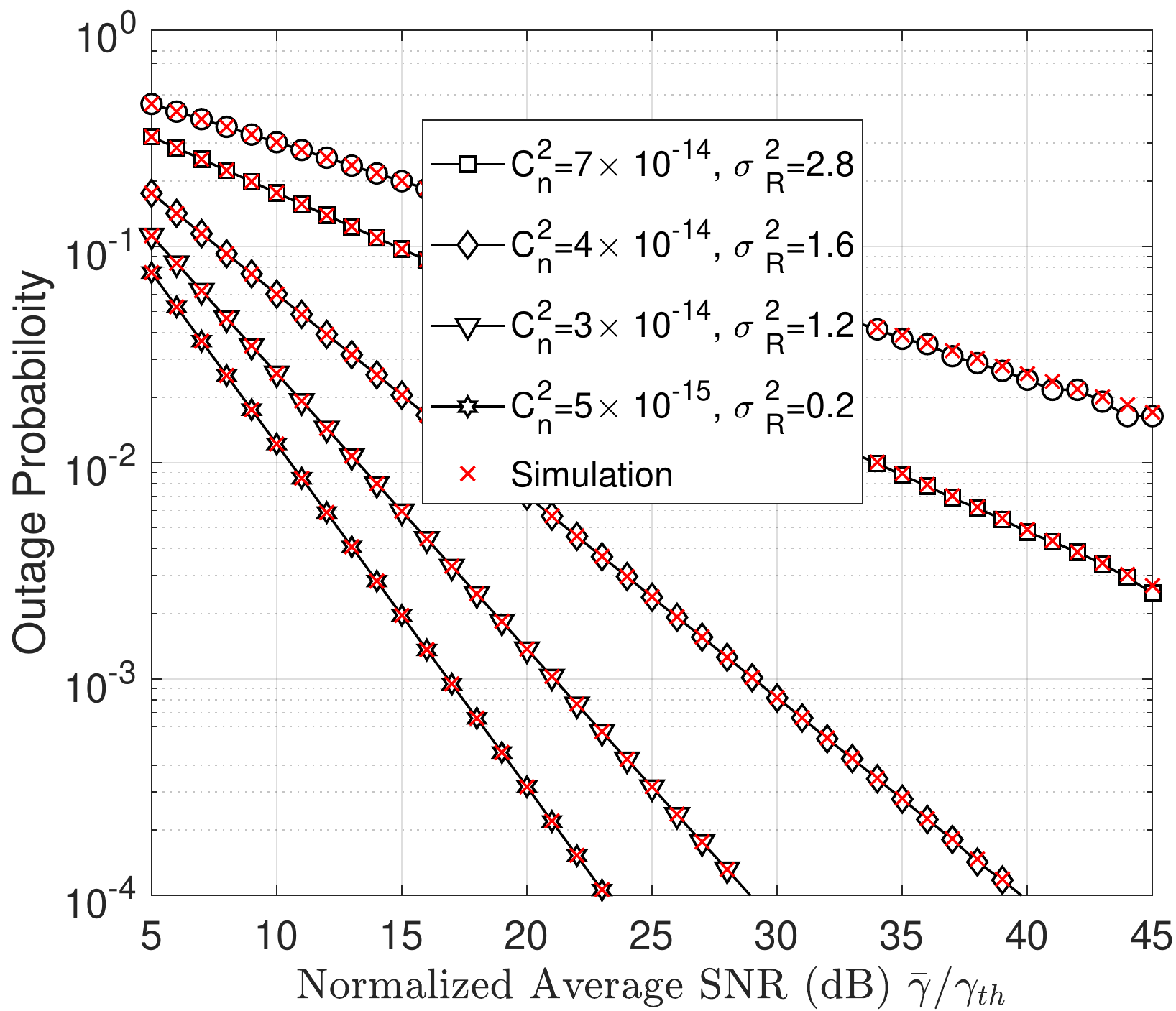}
		\caption{}
	\end{subfigure}~
	\begin{subfigure}[b]{0.45\columnwidth} 
		\includegraphics[width=\columnwidth, height=6.5cm]{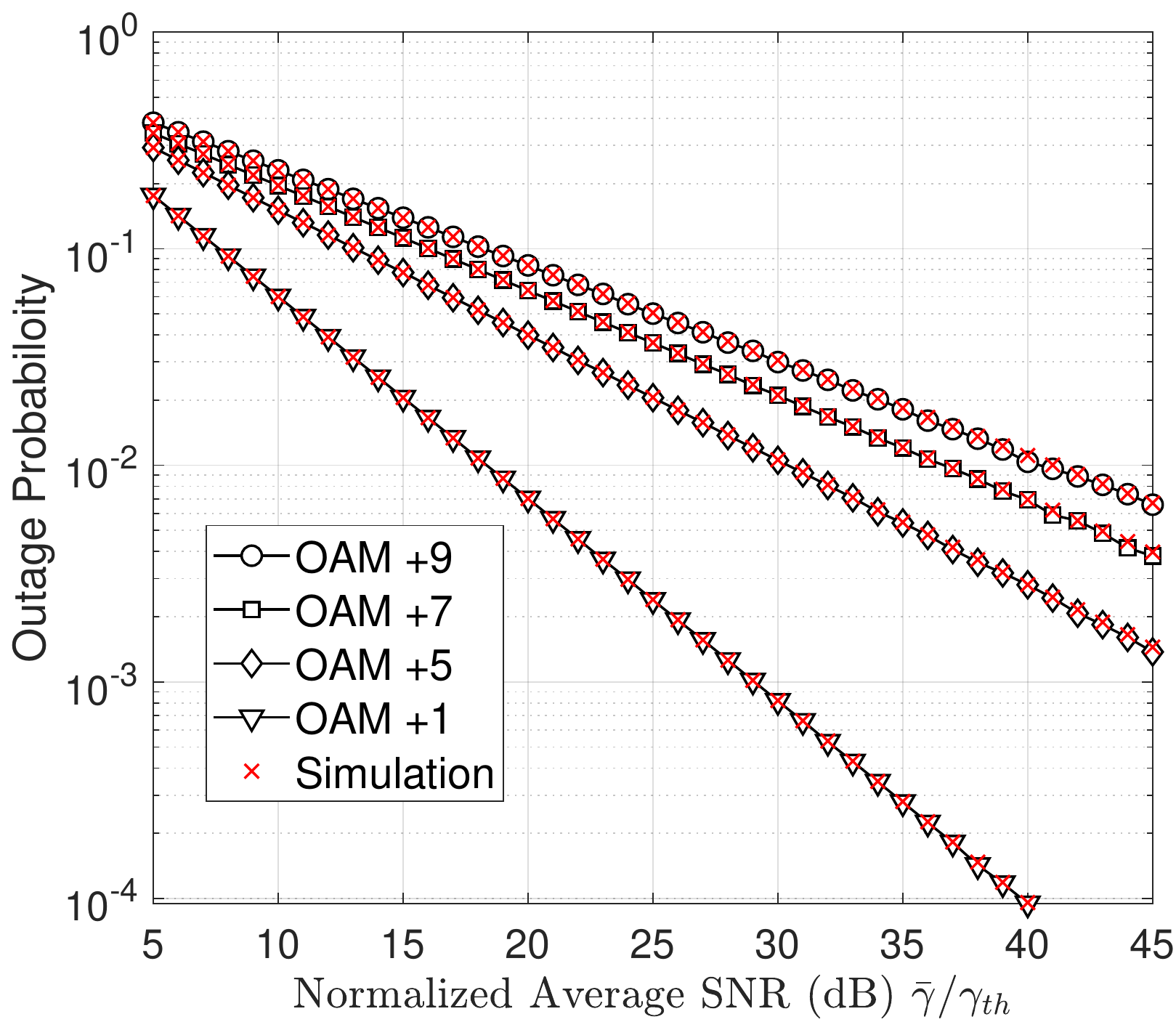}
\caption{}	
\end{subfigure}
	\caption{(a): Outage probability vs  SNR for OAM mode $m=+1$ for different levels of atmospheric turbulence. (b): Outage probability vs the SNR for different OAM modes with $C^2_n=4\times 10^{-14}$ and $\sigma _R^2=1.60$.}
	\label{OutageCap}
\end{figure}

\begin{figure}[h]
	\centering
	\begin{subfigure}[b]{0.45\columnwidth} 
		\centering
		\includegraphics[width=\columnwidth, height=6.5cm]{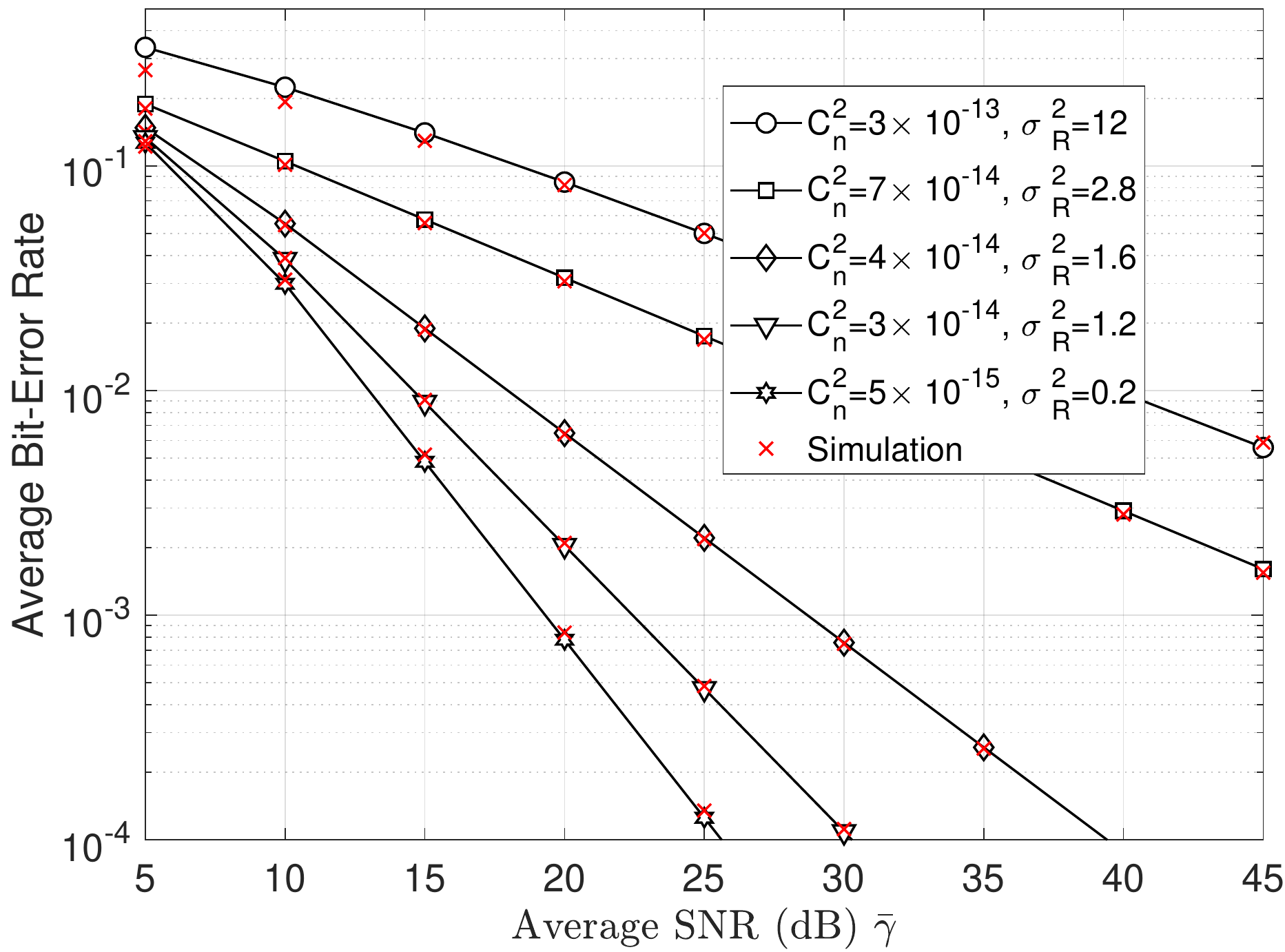}
		\caption{}
	\end{subfigure}~
	\begin{subfigure}[b]{0.45\columnwidth} 
		\includegraphics[width=\columnwidth, height=6.5cm]{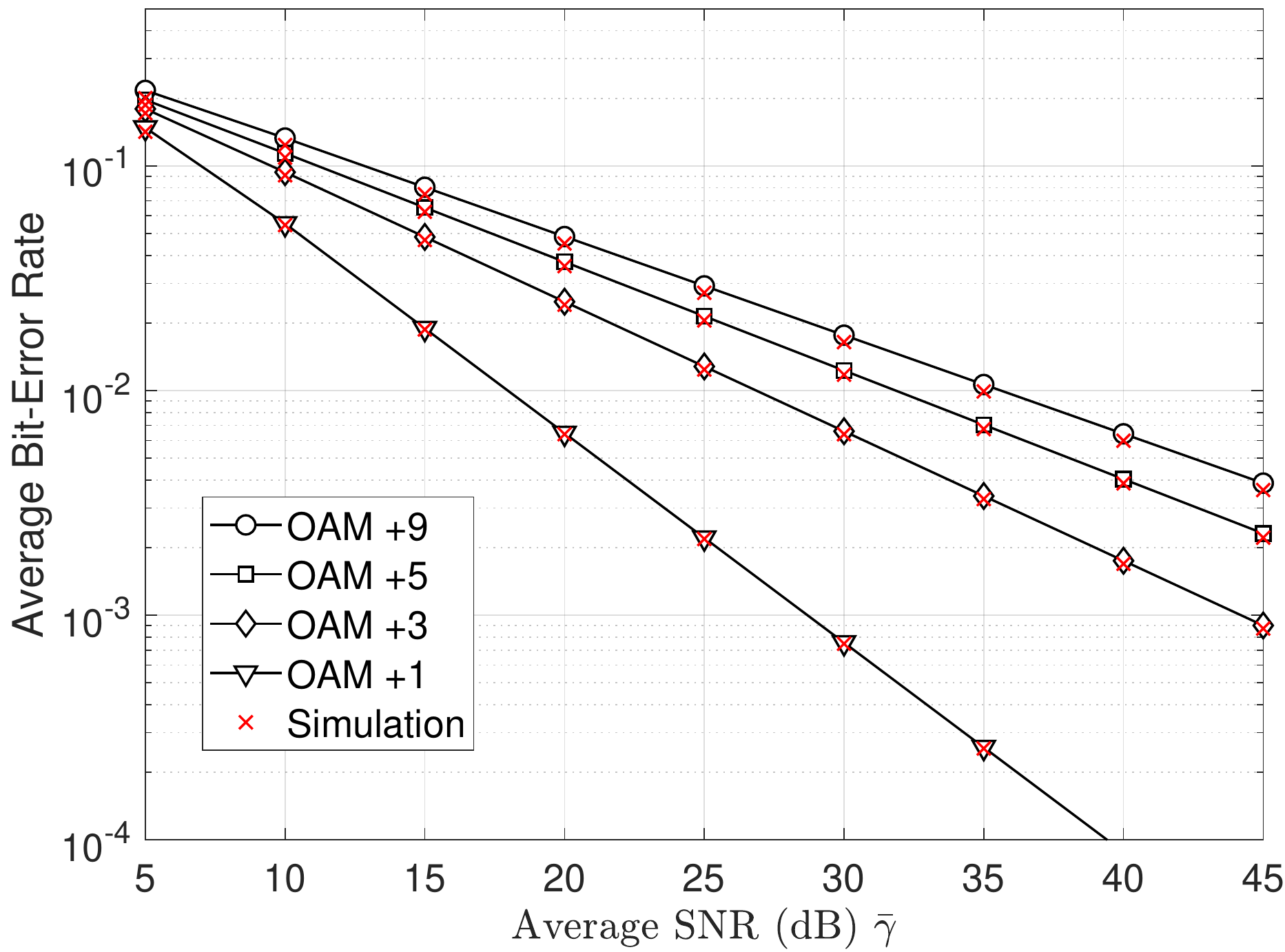}
\caption{text}	
\end{subfigure}	
	\caption{(a): BER  vs SNR for OAM mode $m=+1$ for different levels of atmospheric turbulence. (b): BER  vs SNR for different OAM modes with $C^2_n=4\times 10^{-14}$ and $\sigma _R^2=1.60$.}
	\label{AverageBER}
\end{figure}

In this section, numerical simulations of the previous defined performance metrics are presented and compared to Monte Carlo simulations. In Fig. \ref{CapacitiesErg}(a), we plot the ergodic capacity as a function of the average SNR for the SISO link using OAM mode $m=+1$ for different levels of atmospheric turbulence. From the figure, we notice that the capacity decreases with increasing turbulence. Furthermore, we clearly observe that the analytical capacity curves have a perfect match with the simulated capacities which confirms the accuracy of our derivations. In Fig. \ref{CapacitiesErg}(b), we compare the ergodic capacities using different OAM modes for the moderate to strong turbulence regime given by $C^2_n=4\times 10^{-14}$ and $\sigma _R^2=1.60$. The highest capacity is obtained for OAM mode $m=+1$ followed by other OAM modes having slightly the same capacity. This can be explained by the fact that OAM  mode $m=+1$ is the less affected by atmospheric turbulence and as shown from Fig. \ref{weak_turbulence_dist} the power distribution of OAM mode $m=+1$ remains mostly in the same mode.
\newline
The outage probability of OAM mode $m=+1$ is presented in Fig. \ref{OutageCap}(a) as a function of the normalized average SNR for different levels of atmospheric turbulence. From the figure, we notice that the outage probability increases with increasing atmospheric turbulence which degrades the performance. For example, at a normalized average SNR of 20 dB, when the Rytov variance is $\sigma _R^2=0.2$, $\text{P}_{out}=3\times 10^{-4}$ and for $\sigma _R^2=1.2$, $\text{P}_{out}=1.5\times 10^{-3}$. In Fig. \ref{OutageCap}(b), we compare the outage probabilities obtained by different OAM modes for a moderate to strong atmospheric turbulence regime with $C^2_n=4\times 10^{-14}$ and $\sigma _R^2=1.60$. From the figure, we notice that the lowest outage performance is reached by OAM mode $m=+1$ and as the OAM mode order increases the outage performance degrades.
\newline
In Fig. \ref{AverageBER}(a) and \ref{AverageBER}(b), the average BER is plotted as a function of the average SNR for different OAM modes and at several levels of atmospheric turbulence. From the figures, we can notice an excellent match between theoretical results and simulations, and as for the outage probability, OAM mode $m=+1$ achieves the lowest BER performance. Consequently, from all the previously obtained results, the proposed analytical performance metrics derived from the GGD are very accurate and  perfectly match numerical simulations.

\section{MIMO OAM Links with Spatial Diversity}
\begin{figure}
	\centering
	\begin{subfigure}[b]{0.5\columnwidth} 
		\centering
		\includegraphics[width=\columnwidth, height=6.5cm]{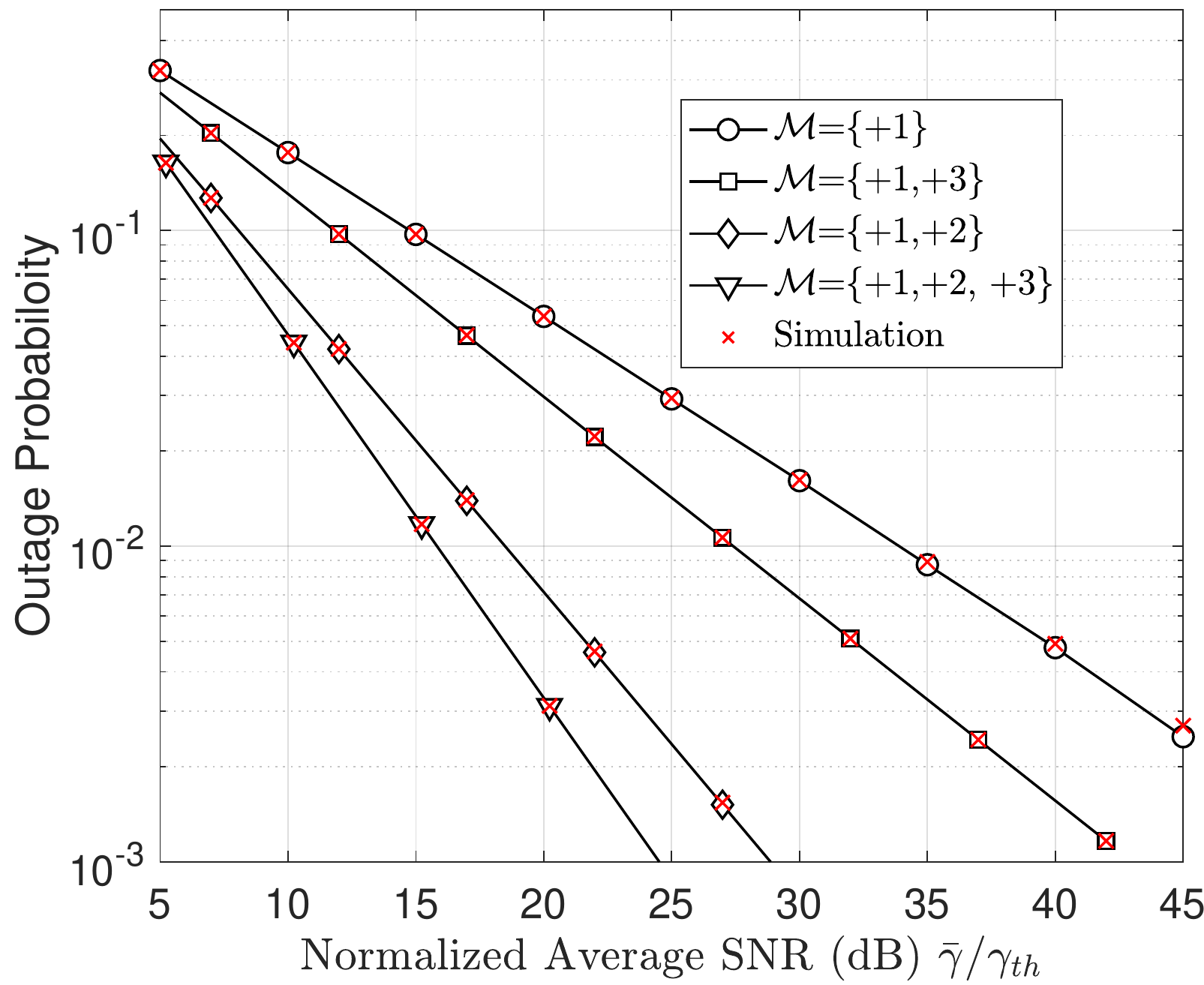}
		\caption{}
	\end{subfigure}~
	\begin{subfigure}[b]{0.5\columnwidth} 
		\includegraphics[width=\columnwidth, height=6.5cm]{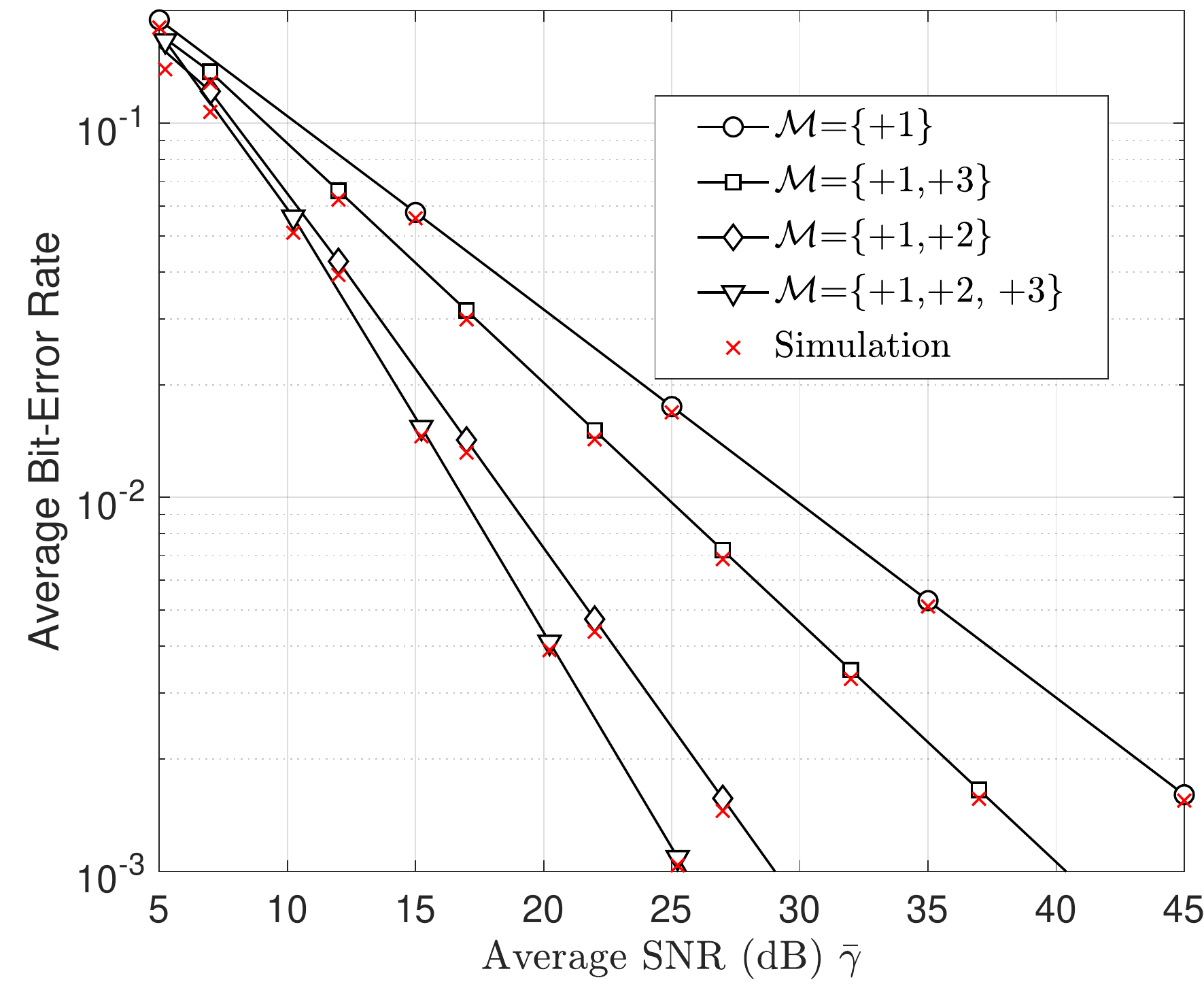}
		\caption{}
	\end{subfigure}
	\caption{ Performance of MIMO OAM FSO system for $C^2_n=7\times 10^{-14}$ and  $\sigma _R^2=2.80$. (a): Outage probability  vs normalized SNR.  (b): BER  vs average SNR. }
	\label{MIMOPerformance}
\end{figure}

In this section, we propose to enhance the error performance of OAM FSO systems by using  multiple OAM beams to transmit the same data. Typically, FSO systems have used multiple transmitters and receivers having apertures that are separated enough  to bring diversity gain \cite{navidpour2007ber,khalighi2009fading}. However, this strategy comes at the cost of more hardware complexity and  apertures spacing limitations which cannot always be satisfied. Spatial diversity can be used at the receiver in the form of a single-input multiple-output (SIMO) system as in \cite{Huang}. At the receiver, combining techniques such as equal gain combining (EGC), selective combining (SC) or maximum ratio combining (MRC) can be used. Spatial diversity can also be set at the transmitter in the form of a multiple-input single-output (MISO) system \cite{tsiftsis2009optical}. In this case, beamforming can be used   thanks to channel state information to improve the performance \cite{huang2017spatial}. Moreover OAM mode selection can also be used to achieve better rates  and lower BER \cite{amhoud2018oam,Huang}. In the absence of channel state information, space-time coding can also be used as an alternative solution to increase the performance \cite{amhoud2018oam}. Spatial diversity can be used at both the transmitter and the receiver in the form of a MIMO system \cite{tsiftsis2009optical}. 
We consider a square MIMO system $\left( \mathcal{M}=\mathcal{N}\right) $  where $M=\left | \mathcal{M}  \right |$  OAM modes are transmitted and received. By using a MRC at the receiver, the electrical SNR is given by \cite[Eq. (14)]{tsiftsis2009optical}:
\begin{eqnarray}
\gamma_{\text{M}}=\frac{\eta^2}{M^2N_0} \sum_{n \in \mathcal{N}}^{}\left ( \sum_{m \in \mathcal{M}}^{} I_{mn}\right )^2,
\label{GammaMIMO}
\end{eqnarray}
where the factor $1/M^2$  in the previous equation allows to compare the SISO and MIMO systems at the same transmitted optical power.
\newline
The derivation of the exact PDF of $\gamma_{\text{M}}$ is complex as it involves sums of GGDs that are non identically distributed. To overcome this issue and compare the performance of SISO and MIMO configurations, we approximate $I_{\text{M}}=\frac{1}{M} \sqrt{\sum\limits_{n \in \mathcal{N}}^{}\left ( \sum\limits_{m \in \mathcal{M}}^{} I_{mn}\right )^2}$ by a GGD and estimate its parameters.  Therefore, the PDF and CDF of $\gamma_{\text{M}}=\left ( \eta I_{\text{M}} \right )^2/N_0$ can be directly given from Eq. (\ref{PDF_SNR}) and Eq. (\ref{CDF_SNR}), and the outage probability and average BER can be theoretically computed using Eq. (\ref{Pout}) and Eq. (\ref{BER}).
\newline
Before performance evaluation of the MIMO system, it is important to choose the appropriate OAM modes in order to fully exploit the benefit of the spatial diversity. When  an OAM mode of order $m$ is used for transmission, it is normal to detect the power conserved in that mode first. In fact, the power $I_{mm}$ is much higher than the power spread into other modes given by $I_{mn}$ as shown from Figs. \ref{weak_turbulence_dist}, \ref{moderate_turbulence_dist}, and \ref{saturation_turbulence_dist}. 
When using spatial diversity, the detected signal on OAM mode $n$ contains the power transmitted from OAM mode $n$ and also crosstalk powers from other transmitted OAM modes as given by Eq. (\ref{MIMO}). 
Furthermore, the performance achieved by the spatial diversity is limited by the correlation between  spatial paths \cite{navidpour2007ber}. Therefore, to achieve the maximum spatial diversity gain,  the correlation between the self-channel irradiance $I_{mm}$ and the crosstalk  $I_{mn}$ has to be minimal. To have an insight on the correlation between OAM modes, we propose to compute the correlation coefficient between $I_{mm}$ and $I_{mn}$  given by:
\begin{equation}
\rho_{mn}=\frac{\text{cov}\left ( I_{mm},I_{mn} \right )}{\sigma_{I_{mm}} \sigma_{I_{mn}}},
\label{corrcoef}
\end{equation}
where $\text{cov}\left ( \cdot ,\cdot  \right )$ is the covariance and $\sigma_{ (\cdot) }$ is the standard deviation.
\begin{figure}[H]
	\centering
	\includegraphics[width=8cm, height=7cm]{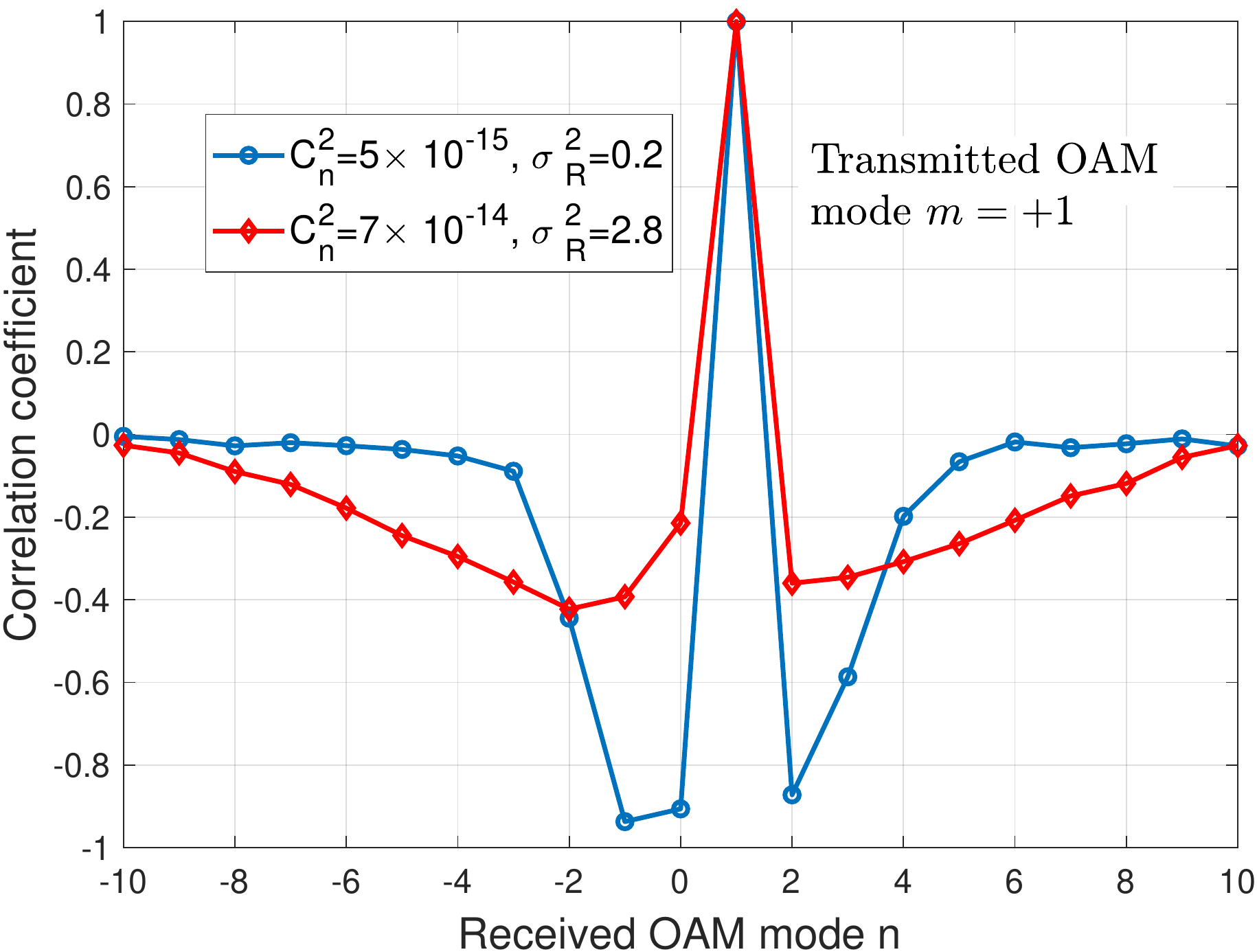}
	\caption{The correlation coefficient between  $I_{+1+1}$ and $I_{+1n}$ for different atmospheric turbulence regimes.}
	\label{CorrCoeff}
\end{figure}
Consider that OAM mode $m=+1$ is transmitted,  we compute the correlation coefficients between $I_{+1+1}$ and $I_{+1n}$ for $n\in \mathcal{N}=\left \{ -10,-9,...,+9,+10 \right \}$. In Fig. \ref{CorrCoeff}, we plot the correlation coefficients for the weak (resp. strong) turbulence regimes given by $C^2_n=5\times 10^{-15}$ and $\sigma _R^2=0.2$ (resp. $C^2_n=7\times 10^{-14}$ and $\sigma _R^2=2.8$). From Fig. \ref{CorrCoeff}, we notice that adjacent modes to the transmitted  mode $m=+1$ have inverse correlation coefficient. The latter increases as the mode order gets further from $m=+1$.  Therefore, closer OAM mode orders are preferred to achieve a maximum diversity \cite{Huang}. In addition to the correlation, the power spread from the transmitted mode due to atmospheric turbulence is more important for adjacent OAM modes than further modes. Therefore, higher SNR gain will be provided by using  modes with closer OAM order.
\newline 
In Fig. \ref{MIMOPerformance}, the performance of  OAM FSO system using spatial diversity is presented for the moderate to strong turbulence regime given by  $C^2_n=7\times 10^{-14}$ and $\sigma _R^2=2.80$. For both the outage probability and the average BER, we notice that the set $\mathcal{M}  =\left \{ +1, +2 \right \}$ achieves better performance than the set $\mathcal{M}  =\left \{ +1, +3 \right \}$ as expected. Furthermore, increasing the number of OAM modes  enhances the performance as shown from the curve with the set $\mathcal{M}  =\left \{ +1, +2, +3 \right \}$. In addition, we have also  approximated $I_{\text{M}}$ with a GGD and estimated its parameters which allowed to compute the theoretical outage probability and average BER. The plotted curves based on the GGD approximation give an excellent match with the simulated data. 
\section{Space-Time Coding for OAM FSO Systems}
In the previous section, we have shown that spatial diversity can improve the error performance by sending multiple replicas of the same signal on different OAM modes.  
 Nonetheless, the achieved diversity performance was obtained at the cost of a reduced multiplexing gain. To achieve full diversity and multiplexing gains, we consider space-time coding. ST coding was originally  designed for wireless radio-frequency (RF) systems \cite{tarokh}, and recently investigated for few-mode fibers \cite{amhoud,amhoud2016ICT}. The Golden code \cite{Belfiore} was designed for $2\times 2$ wireless MIMO systems and achieves a full diversity with the best coding gain for Rayleigh fading channels. The Alamouti code \cite{alamouti}, is a half-rate orthogonal ST code that also achieves a full diversity and has the benefit of a very low  decoding complexity that reduces to a simple channel inversion.  ST coding was also demonstrated with heterodyne multiple apertures FSO systems \cite{haas2002space}. However, due to the high cost of heterodyne implementation and complex signal processing required for ST decoding, it is not easily deployed in practice. 
\par 
In addition, ST coding was also investigated for multiple apertures IM/DD FSO systems. In \cite{bayaki2010space}, a theoretical error probability was derived, and ST coding was shown to achieve full diversity. Furthermore, modified versions of known ST codes from RF were constructed and adapted to IM/DD systems. In \cite{safari2008we,simon2005alamouti}  modified versions of the Alamouti code were proposed. The new constructed Alamouti code preserves all the useful properties of the original  code. The codeword matrix of the Alamouti code is given by \cite{safari2008we}:

\begin{equation}
\mathbf{X}_{\text{Al}}=\begin{bmatrix}
s_1 & s_2\\ 
\sim s_2 & s_1
\end{bmatrix},
\end{equation}
where the symbol $\sim s_i$ refers to the bit-wise not of the transmitted bit $s_i$. The main drawback of the Alamouti code is its half-rate code that only allows sending two bits $s_1, s_2$ during two time slots and using two OAM modes.
\newline
Furthermore, Mroueh has proposed the Golden-Light (GL) code \cite{mroueh2019extended} which is an extended version of the Golden code that was adapted to IM/DD  systems. The proposed code was shown to  conserve an optimal and non-vanishing determinant. The  GL codeword  is given by \cite{mroueh2019extended}:
\begin{equation}
\mathbf{X}_{\text{GL}}=\begin{bmatrix}
\frac{1}{\sqrt{1+ z^{*2}_1}}\left ( s_1-s_2z^{*}_1 \right ) & \frac{1}{\sqrt{1+ \bar{z}^{*2}_2}}\left ( s_3+s_4\bar{z}^{*}_2 \right )\\ 
\frac{1}{\sqrt{1+ z^{*2}_2}}\left ( s_3-s_4z^{*}_2 \right ) & \frac{1}{\sqrt{1+ \bar{z}^{*2}_1}}\left ( s_1+s_2 \bar{z}^{*}_1 \right )
\end{bmatrix},
\end{equation}
where $z^{*}_1=\frac{1}{2}\left ( 1+\sqrt{5} \right )$ and $z^{*}_2= 2+\sqrt{5}$ are specific algebraic numbers related to the code construction. In addition to a  full diversity, The GL code has a full-rate, which means that during two time slots and using two OAM modes, the GL code allows transmitting four bits $s_1, s_2, s_3, s_4$.
\newline
To evaluate the performance of ST coding in OAM FSO systems affected by atmospheric turbulence, we consider that $M$ OAM modes are transmitted and detected.  The resulting  MIMO transmission system  is given by:
\begin{eqnarray}
\mathbf{Y}= \eta \mathbf{I}\mathbf{X}+\mathbf{N},
\label{eqSTcodes}
\end{eqnarray} 
where  $\mathbf{X}\in \mathbb{R}^{M \times M}$  and $\mathbf{Y}\in \mathbb{R}^{M \times M}$ are the transmitted and received codewords. $\mathbf{N}\in \mathbb{R}^{M \times M}$ represents an additive white Gaussian noise with zero mean and a variance $N_0$ per complex dimension. $\mathbf{I}$ represents the channel matrix with entries $I_{mn}$. In the next section, we derive a theoretical expression of the error probability of ST coded OAM FSO systems. 
\subsection{Derivation of the Error Probability}
Let  $\mathbf{X}_i$   be the transmitted  codeword and $\mathbf{X}_j$ the estimated codeword. Considering a maximum-likelihood decoder, the error probability is defined as:
\begin{equation}
P_{\text{e}}=\sum_{\mathbf{X}_i\in \mathfrak{C}}{} \text{P}_\text{r} \left \{ \mathbf{X}_i \right \}\text{P}_\text{r} \left \{ \mathbf{X}_j\neq  \mathbf{X}_i  \right \}.
\end{equation}
For equiprobable codewords and using the union bound of the error probability \cite{Proakis}, we obtain:
\begin{equation}
P_{\text{e}} \leq \frac{1}{\left | \mathfrak{C} \right |} \sum_{\mathbf{X}_i \neq  \mathbf{X}_j} \text{P}_\text{r} (\mathbf{X}_i, \mathbf{X}_j),
\end{equation}
where  $\mathfrak{C}$ is the set of all possible codewords and $\ \text{P}_\text{r} (\mathbf{X}_i, \mathbf{X}_j)$ is the average pairwise error probability (PEP) of  transmitting $\mathbf{X}_i$ and decoding $\mathbf{X}_j$. Let  $\mathbf{X}=\mathbf{X}_i-\mathbf{X}_j$ denotes the difference of two codewords, the evaluation of the error probability simplifies to the averaging of the conditional PEP given by \cite{bayaki2010space}: 
\begin{equation}
\text{P}_\text{r} (\mathbf{X}_i, \mathbf{X}_j/\mathbf{I})=Q\left( \sqrt{\frac{\eta^2}{N_0 M^2 }\left \| \mathbf{IX} \right \|^2}\right) 
\label{eq.pep}
\end{equation}
\begin{equation*}
\text{with}~~~~Q(x)=\frac{1}{\pi}\int_{0}^{\pi /2}\exp\left ( -\frac{x^2}{2\sin^2\theta} \right )d\theta
\end{equation*}
\begin{align}
\text{and}~~~\left \| \mathbf{IX} \right \|^2&=\text{tr}\left ( \mathbf{IXX^{\ast} I^{\ast}}  \right )\\
&=\sum_{n=1}^{M}\mathbf{I}_n\mathbf{X^\Delta} \mathbf{I}_n^{\ast},
\end{align}
where $\mathbf{I}_n=(I_{1,n},...,I_{M,n})$. $\mathbf{X^\Delta=XX^{\ast}}$ is a hermitian matrix, thus there exists a unitary matrix $\mathbf{U}$ and a real diagonal matrix $\mathbf{D=\text{diag}\left (  \lambda _1,..., \lambda _M \right )}$ such that $\mathbf{X^\Delta=UDU^{\ast}}$,  hence:
\begin{equation}
\left \| \mathbf{IX} \right \|^2=\sum_{n,m=1}^{M}\lambda _m  \beta _{mn} ^2
\label{eq.norm},
\end{equation}
with $\beta _{mn}= \sum_{k=1}^M u_{km}I_{kn}$. By injecting the previous equation in Eq. (\ref{eq.pep})  and using the definition of the $Q$ function, the pairwise error probability becomes:
\begin{equation}
\text{P}_\text{r} (\mathbf{X}_i, \mathbf{X}_j/\mathbf{I} )
=\frac{1}{\pi }\int_{0}^{\pi/2}\exp\left (-\sum_{m,n=1}^{M} \frac{  \beta _{mn}^2 \zeta _m}{\sin^2 \theta } \right )d \theta,
\label{FInalPEP}
\end{equation}
 where we have denoted $\zeta  _m=\frac{\eta^2\lambda_m}{2N_0 M^2} $. 
 \newline
A further derivation of the previous equation requires the knowledge about the distribution  of  $\beta _{mn}$ which is not accessible due to the terms $ u_{km}$.
However, in the case of orthogonal ST codes we have  $\mathbf{X}^\Delta=z^\Delta\mathbbm{1} _2$ \cite{tirkkonen2002square}, where $z^\Delta=\sum_{k}\left | x_{km}^\Delta -x_{kn}^\Delta \right |^2$, and $x_{mn}^\Delta$ are the elements of the matrix $\mathbf{X}^\Delta$ and $\mathbbm{1} _2$ is the two dimensional identity matrix. Therefore in the case of orthogonal ST codes,  Eq. (\ref{FInalPEP}) becomes:
\begin{equation}
\text{P}_\text{r} (\mathbf{X}_i \perp \mathbf{X}_j/\mathbf{I} )
=\frac{1}{\pi }\int_{0}^{\pi/2}\exp\left (-\frac{\eta^2 z^\Delta }{2N_0 M^2  \sin^2 \theta}\sum_{m,n=1}^{M}   I _{mn}^2  \right )d \theta,
\label{FInalPEPOrtho}
\end{equation}
As detailed in the previous section, the  fading coefficients $I_{mn}$ are not independent. Hence, it is not possible to apply the approach used in \cite{tarokh} that consists in taking the average of the conditional pairwise error probability and writing the moment of the product as the product of moments.
To overcome this, we approximate  the random variable $\sum_{m,n=1}^{M}  I _{mn} ^2 $ with a GGD. Therefore, the SNR of orthogonal ST codes can be written as $\gamma_{\perp }= \frac{\eta^2}{N_0 M^2} \sum_{m,n=1}^{M}  I _{mn} ^2$. Hence, the average pairwise error probability of orthogonal ST codes becomes:
 \begin{equation}
 \text{P}_\text{r} (\mathbf{X}_i \perp \mathbf{X}_j)
 =\frac{1}{\pi }\int_{0}^{\pi/2}\textup{M}_{\gamma_{\perp }}\left ( \frac{z^\Delta}{2\sin^2 \theta} \right )d\theta,
 \label{FInalPEPOrtho1}
 \end{equation}
where $\textup{M}_{\gamma_{\perp }}\left ( \cdot  \right )$ is given in equation  (\ref{MGF}) or (\ref{MGF2}).
\subsection{Numerical Results}
We consider a $2 \times 2$ MIMO system using the set of OAM modes $\mathcal{M}  =\left \{ +1, +2 \right \}$,  and we compare the performance obtained by using ST coding to the  spatial diversity. To make a fair comparison between the different schemes, we compare the BER performances at the same transmitted bits per channel use (bpcu). For this purpose, we use an L-aray pulse amplitude modulation (PAM). In the case of the GL code, it is possible to send 4 modulated symbols during 2 time slots. We use a 2-PAM modulation to obtain a rate of 2 bpcu. To satisfy the same rate, we use a 4-PAM modulation in the case of the Alamouti code and the spatial diversity scheme.
\begin{figure}
	\centering
	\includegraphics[width=8cm, height=7cm]{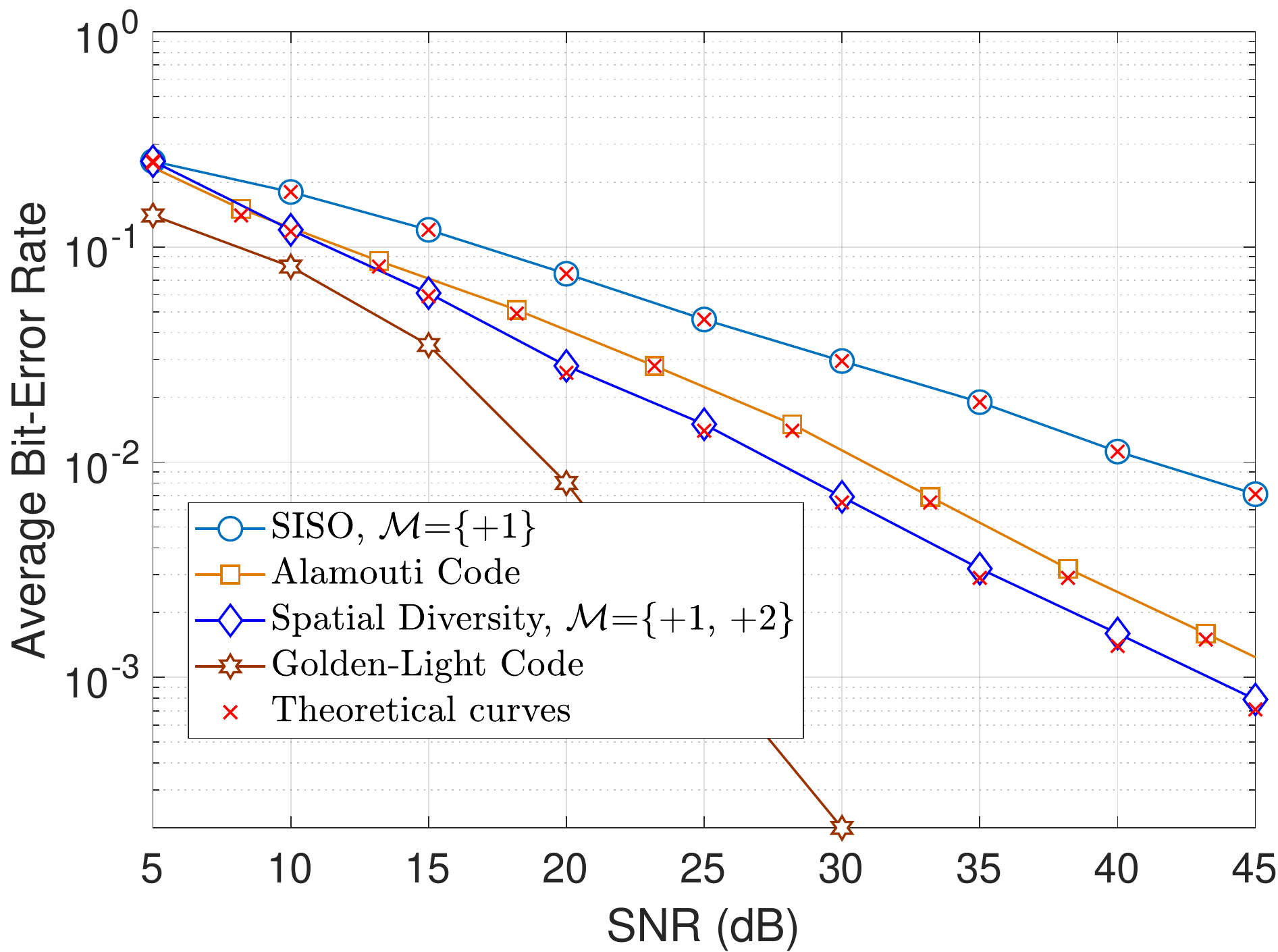}
	\caption{BER performance comparison between ST coding schemes and spatial diversity.}
	\label{STResults}
\end{figure}
In Fig. \ref{STResults}, we plot the obtained BER performance. The SISO link using OAM mode $m=+1$  is also plotted as a reference. From the figure, we notice that the GL code outperforms all other schemes. This is due to the coding gain of the GL code and also the use of the 2-PAM modulation instead of the 4-PAM for other schemes. Moreover, we notice that the spatial diversity scheme outperforms the Alamouti code which is consistent with the previous results obtained in \cite{safari2008we,sapenov2018diversity}. In addition, the theoretical performance of the Alamouti code based on  the GGD approximation has a good match with the simulated performance. 
\section{Conclusion}
In this paper, we have proposed a new statistical model for atmospheric turbulence fading in OAM FSO transmission systems. The proposed GGD can model the self-channel irradiance of OAM modes as well as the crosstalk between OAM modes for all atmospheric turbulence regimes. Hence, the proposed  model can be used  to overcome the computationally complex Monte Carlo simulations to simulate the propagation of OAM beams. Afterward, we have derived analytical expressions for the ergodic capacity, the outage probability, and the average BER. The obtained results show  perfect match with  simulations. Moreover, we have proposed to enhance the performance of OAM FSO systems by considering spatial diversity where several OAM modes are used to transmit the same data. From the obtained outage probability and average BER, significant performance improvements were achieved. Finally, ST coding was proposed to achieve  full diversity and multiplexing gains. A theoretical derivation of the error probability was conducted, and numerical simulations of different ST codes were presented.

\bibliographystyle{ieeebib}
\bibliography{Bibliography}

\end{document}